\documentclass[]{article}
\usepackage[a4paper, total={6in, 8in}]{geometry}
\usepackage{authblk}
\usepackage{natbib}
\usepackage{graphicx}
 \usepackage{amsmath}

\title{Precise Physical Parameters, Habitability, and Orbital Stability of Sun-like SB2 Systems: HD 130669, HD 184467, HD 191854, and HD 214222}
\author{Ahmad Abushattal$^1$, Nikolaos Georgakarakos$^{2,3}$, Mashhoor A. Al-Wardat$^4$,\\
Bilal Algnamat$^1$, Hassan B. Haboubi$^4$, Deshinta Arrova Dewi$^5$, \\Enas M. Abu-Alrob$^6$, and Abdallah M. Hussein$^6$}

\affil[1]{Department of Physics, Al-Hussein Bin Talal University, PO Box 20, 71111, Ma'an, Jordan}
\affil[2]{Division of Science, New York University Abu Dhabi, Abu Dhabi, UAE}
\affil[3]{Center for Astrophysics and Space Science (CASS), New York University, Abu Dhabi, PO Box 129188, Abu Dhabi, UAE}
\affil[4]{Department of Applied Physics and Astronomy, and Sharjah Academy for Astronomy, Space Sciences and Technology, University of Sharjah, PO Box 27272, Sharjah, UAE}
\affil[5]{Faculty of Data Science and Information Technology, INTI International University, Malaysia}
\affil[6]{Department of Physics, Faculty of Sciences, Al al-Bayt University, Mafraq, Jordan}

\begin{document}

\maketitle

\begin{abstract}
This work analyzes four Sun-like double-lined spectroscopic binary (SB2) systems by combining visual and spectroscopic observational data with Al-Wardat’s atmospheric modeling method to accurately determine their fundamental parameters. For each system, we determine stellar masses, orbital parallaxes, effective temperatures, spectral types, semimajor axes, and eccentricities with high precision, resolving discrepancies between astrometric and spectroscopic measurements. Moreover, we assess the potential for stable planetary orbits in these systems. We also calculate habitable zones around these binaries based on the orbital evolution of planetary orbits. These systems may represent promising targets for future extrasolar planet searches around Sun-like stars due to their robust physical and orbital parameters that can be used to determine planetary habitability and stability.
\end{abstract}

{\it Unified Astronomy Thesaurus concepts}: Spectroscopic binary stars (1557); Habitable zone (696); Visual binary
stars (1777)

\section{Introduction}
\subsection{Importance and Prevalence of Binary Systems}

The coming generations of telescopes and space missions will lead to discoveries that will answer many questions about the
formation and dynamic evolution processes of binary and
multiple stars. They will also prove helpful in searching for
stellar and planetary companions of those stars, especially Earth-
like ones. Statistically, it has become rather clear that most of the
nearby stars are members of binary and multiple stellar systems,
which form in the early stages of star formation (A. Duquennoy
and M. Mayor 1991; D. Raghavan et al. 2010; G. Duchene and
A. Kraus 2013; D. Jack et al. 2020; B. Algnamat et al. 2022;
H. Alameryeen et al. 2022; S. Alnaimat et al. 2025). Multiple-star
system analysis is an active field of research aimed at
understanding the origin and evolution of stellar systems
(E. P. Horch et al. 2009; G. Duchene and A. Kraus 2013;
A. Tokovinin 2014; A. Tokovinin and E. P. Horch 2016;
J. A. Docobo et al. 2017; R. A. Matson et al. 2018; A. Taani
et al. 2019; A. Tokovinin et al. 2019; L. Piccotti et al. 2020). We
can calculate the telescope size required to visually resolve
spectroscopic binaries by estimating the angular size of the
photocentric orbit (S. Ren and Y. Fu 2010; A. A. M. Abushattal
2017; A. A. Abushattal et al. 2019b).

\subsection{Astrometric Missions and Data Importance}

Hipparcos and Gaia, as European Space Agency (ESA) space-astrometric missions, are considered some of the most essential
achievements in the space sciences over the past decades. They
constitute the cornerstone and a significant source of information
covering various astronomical fields. These missions give the
positions, color indices, motion, distances, and other data for
millions of stars (M. Perryman et al. 1997; F. van Leeuwen 2007;
T. Prusti et al. 2016; Gaia Collaboration 2018a). The parallax is
one of the most essential pieces of information these space
missions provide. It is an important factor in the analysis of
multiple and binary stars. Based on the well-known Kepler’s third
law, the catalogs of these missions use the parallaxes to estimate
the total mass of visual binaries.

\subsection{Discrepancies in Parallax and Complementary Methods}

In recent years, there have been discrepancies between the
parallaxes for the respective space missions Hipparcos and Gaia
(see Table 1). The accuracy of these parallaxes, together with
accurate measurements of the orbital elements, period, and
semimajor axis, will help calculate the most accurate total masses
for the binaries (J. C. Zinn et al. 2017; R. Drimmel et al. 2019;
E. Vasiliev 2019; M. A. Al-Wardat et al. 2021a; M. A. Fardal
et al. 2021). In recent years, hundreds of research papers have
been published in the field of binary stars using many
astronomical methods to calculate the orbital and physical
parameters of those stars, such as, for example, the method of
Tokovinin, the Docobo geometrical methods, and the Al-wardat
method as a complementary atmospheric modeling method
(J. Docobo et al. 1988; A. Tokovinin 1992a, 1992b; J. Docobo
et al. 2000; M. A. Al-Wardat 2007, 2009).

\begin{table*}[h]
\centering
\scriptsize
\caption{Parameters of systems.} \label{parameter}
\begin{tabular}{ccccccc}
\hline\hline
Property & HD 130669 & HD 184467  & HD 191854 &  HD 214222 & Source of data \\  \hline \hline
$\alpha_{2000}$  & $14^{h} 49^{m} 13^{s}.62$ & $19^{h} 31^{m} 07^{s}.96$ & $20^{h} 10^{m} 13^{s}.32$ & $22^{h} 35^{m}42^{s}.67$ & (M. Wenger et al. 2000)   \\
$\delta_{2000}$ & $+10^{\circ}12^{\prime}52^{\prime\prime}.06$ & $+58^{\circ}35^{\prime}09^{\prime\prime}.60$ & $+43^{\circ}56^{\prime}44^{\prime\prime}.07$ &  $+53^{\circ}12^{\prime}14^{\prime\prime}.41$ & (M. Wenger et al. 2000) \\
$m_V$ {[}mag{]}   & 8.42  & 6.59  & 7.606  &  8.036 & (Esa 1997) \\
$\pi_{Hip 1}$ {[}mas{]}  & - & $59.84\pm 0.64$  & $19.45\pm 0.80$ & $13.76\pm0.79$  & (F. van Leeuwen 2007) \\
$\pi_{Hip 2}$ {[}mas{]}  & $22.59\pm1.23$  & $58.96\pm 0.65$ & $19.48\pm 0.54$  & $ 14.15\pm 0.74$ & (M. Perryman et al. 1997) \\
$\pi_{Gaia 1}$ {[}mas{]}  & - & $58.37\pm 0.54$ & - & $46.61\pm 0.83$  & (A.G. Brown et al. 2016) \\
$\pi_{Gaia 2}$ {[}mas{]}  & -   & $54.8715\pm 0.24$  & $19.2965\pm 0.1282$ & $14.235\pm 0.319$ & (Gaia Collaboration et al. 2018b)  \\
$\pi_{Gaia 3}$ {[}mas{]}   & - & $52.2767\pm 0.3519$  & -  & -  &  (Gaia Collaboration et al. 2021) \\ \hline\hline\
\end{tabular}
	\\
	$^*$ http://simbad.u-strasbg.fr/simbad/sim-fid.
\end{table*}

\subsection{Observational Techniques and Types of Binaries}
Astrometric, spectroscopic, and eclipsing binaries are the main
types of binary systems. It is challenging to determine the 
individual masses of a system using a single technique. Fifty
years ago, speckle interferometry entered the field of observing
binary stars effectively and accurately, and it has become
a significant part of the observational processes employed
by researchers around the world (A. Labeyrie 1970;
H. A. McAlister 1976, 1977, 1978; I. Balega et al. 1984;
D. Bonneau et al. 1986; A. Abushattal et al. 2022a). Two types of
binary systems correlate with the masses of the components of
the binary star: the astrometric binary (AB) and the double-lined
spectroscopic binary (SB2). In AB, the orbit solution is
determined via the parallax, and by applying Kepler’s third
law, we can determine the total mass of the system. In the SB2
orbital solution, we can determine the mass ratio of the
components, and by combining both methods, we can determine
the three-dimensional orbit, the individual masses, and the orbital
parallax. Measuring the orbital parallax is considered an effective
verification method for astrometric parallaxes after systematic
errors were found in the Gaia data release (Gaia DR2) for bright
stars (K. G. Stassun $\&$ G. Torres 2018; R. Drimmel et al. 2019;
E. Vasiliev 2019; A. Abushattal et al. 2022b). During recent
years, there have been many publications that relied on such a
research method to determine the physical parameters with a
combination of each technique (J. A. Docobo et al. 2014, 2018c;
X.-L. Wang et al. 2016; J. A. Docobo et al. 2017; K. V. Lester
et al. 2019a, 2019b, 2020; L. Piccotti et al. 2020). Precision
astrometric interferometry in binary star systems is a promising
technique for planet searches that complements the radial-
velocity method that has been so successful. Internal errors of a
few microarcseconds are possible if the systematic errors can be
controlled (J. T. Armstrong et al. 2004).

\subsection{Binary Systems and Exoplanet Habitability}

Main-sequence and pre-main-sequence stars constitute
approximately $50\%$ of multiple-star systems, positioning
binary systems as favorable environments for exoplanet
exploration by space missions such as CoRoT, Kepler, K2,
and TESS, intending to identify Earth-like planets
(G. R. Ricker et al. 2010; S. Eggl et al. 2012; W. J. Borucki
2016; A. Abushattal et al. 2019; S. B. Howell et al. 2021).
Multiple stars often contain the most massive exoplanets, and
binary stars are among the most important hosts of giant
exoplanets in multiple systems (C. Fontanive and D. Bardalez
Gagliuffi 2021). A couple of important features in the study of
exoplanets are the potential habitability of a planet and its
orbital stability, which both depend on the parameters of the
host binary star. Such parameters are the individual masses,
the semimajor axis, the orbital eccentricity, the luminosities,
and temperatures, and they are the main target of this work
(K.-U. Michel $\&$ M. Mugrauer 2021; E. Pilat-Lohinger and
A. Bazso 2021; N. Georgakarakos and S. Eggl 2019).

The key idea of this work is to combine the high-grade visual
solution with a definitive spectroscopic orbital solution for
selected double-lined spectroscopic candidates to calculate the
orbital parallax and the individual masses. Also, we estimate the
atmospheric and fundamental parameters of the individual
components. First, these results will be compared with other
values resulting from the use of different methods, such as
Edwards’s method (T. Edwards 1976; A. A. M. Abushattal 2017)
or Al-Wardat’s method for analyzing binary and multiple stars
(M. A. Al-Wardat 2002, 2012). Second, we will use these results
to describe the three-dimensional orbits for each system.
Third, these quantities, together with the luminosities (e) and
temperatures (T), will be used to assess the potential of planetary
habitability and orbital stability. Finally, we will use our results
to determine the minimum size of the telescope necessary
to visually observe these systems (A. A. M. Abushattal 2017;
A. A. Abushattal et al. 2019b).

\subsection{Al-Wardat’s Method Description and Validation}

Al-Wardat’s method for analyzing stellar systems is a
computational method that utilizes the observed photometry
and spectral energy distributions (SEDs) of entire stellar
systems to build synthetic SEDs for the individual components
of the systems. These systems are barely resolved using space
telescopes or special techniques on ground-based telescopes.
The method constructs synthetic SEDs for the individual
components or variability phases through an iterative process.
The results are used for estimating the physical parameters of
these component, from which their masses, spectral types,
luminosity classes, and populations are defined. The method
uses local thermodynamical equilibrium plane-parallel model
atmospheres, typically using Kurucz ATLAS 12 models, to
generate the synthetic SED for each component. The method
was successfully used to analyze 10s of binary and multiple
stellar systems, as well as some variable stars (M. Al-Wardat
2002a, 2008, 2009, 2012; M. A. Al-Wardat et al. 2016, 2017a;
S. G. Masda et al. 2018, 2019b; A. M. Hussein et al. 2022;
A. A. Abushattal et al. 2024).

\subsection{Objectives and Structure}

In this study, we combine visual and spectroscopic data to
determine the fundamental physical and orbital parameters of
four Sun-like double-lined spectroscopic binary systems. To
resolve astrometric discrepancies, Al-Wardat’s atmospheric
modeling method is applied to derive precise stellar masses,
orbital parallaxes, and other essential parameters. The habit-
ability and orbital stability of hypothetical planets around these
systems are also assessed. This paper is organized as follows:
Section 2 outlines data sources and selection criteria, Section 3
presents orbital analysis, Section 4 presents atmospheric
modeling techniques, Section 5 discusses mass and parallax
determinations, Section 6 examines habitability and stability
conditions, and Section 7 summarizes our findings and their
implications for the future.

\section{Data Resources and Selection of Candidates}

The binary systems HD 130669, HD 184467, HD 191854,
and HD 214222 were chosen based on stringent criteria to
ensure the robustness of our analysis. First, each system is
classified into a double-lined spectroscopic binary (SB 2) in
the ninth catalog of spectroscopic binary orbits (SB9), which
provides information on mass ratios. Additionally, all systems
feature high-quality visual orbital solutions from the Sixth
Catalog of Visual Binary Star Orbits, which enables combined
spectroscopic and visual analysis. Additionally, we prioritized
systems with well-graded orbits (grade 1 or 2 for visual orbits
and grade 5 for spectroscopic orbits; J. Docobo et al. 2018).
This ensured data reliability. Furthermore, the systems were
selected for their Sun-like characteristics, including F-, G-, or
K-type main-sequence components, which makes them
suitable candidates for the study of planetary habitability.
This work focuses on binary stars that possess both spectroscopic and visual orbits, combining the two types into one known as a spectroscopic-visual binary. In the Ninth Catalog
of Orbits of Spectroscopic Binaries (SB9), we first look for
double-lined spectroscopic binary candidates. These types of
binaries are essential for such studies because they provide an
important piece of information: the mass ratio of the binary
components. The first source, SB9, is a catalog of the orbital
elements of spectroscopic binary stars, accessible online at
https://sb9.astro.ulb.ac.be/. It contains valuable information
about identified selected samples, which can be separated into
basic data (coordinates, spectral types, and apparent magni-
tudes) and identifiers using common catalogs (HD, HIP, BD).
The most important part is the orbital solution, which contains
the orbital elements with errors, derived quantities, and the
grade of the orbits (1: poor, 5: definitive) with references
(D. Pourbaix et al. 2004). The second source is the Sixth
Catalog of Orbits of Visual Binary Stars, which can be
accessed online at http://www.astro.gsu.edu/wds/orb6.html.
It includes 3318 systems with 10,089 orbits graded on a scale
of 1 to 5 (1: definitive, 5: poor). On the website, one can find
an introduction, the Orbit grading method, and a description of
the catalog (orbital elements, ephemerides, notes, references).
Thus we search for SB2s with reliable visual and spectroscopic
orbital solutions. This is achieved by combining data from the
sixth catalog with that from SB9 to obtain the appropriate
primary data for such a study.

\section{Visual and Spectroscopic orbits}
The systems we investigate in the work are as follows:

{\it HD 130669, A 2983, HIP 72479}. In 1955, N. G. Roman
listed HD 130669 in the Catalogue of High-velocity Stars as a
K2V star, with a spectroscopic orbit of $0^{\prime\prime}.0275$ and 
photoelectric magnitudes and a color index of $V=8.43$ and
$B-V=0.88$ (N. G. Roman 1955). O. J. Eggen calculated the
dynamical parallax of this system to be $0^{\prime\prime}.022$, based on a
period of 10 yr, with a spectral type of G9 V. Five years later,
he revised it to be $0^{\prime\prime}.025$, and R. F. Griffin suggested a
spectral type of K0V (O. J. Eggen 1955; R. F. Griffin 2015).
Unfortunately, in the SIMBAD database, this system doesn’t
have a Gaia parallax, while Hipparcos calculates it to be
$0^{\prime\prime}.02259$, with a photometry of $V=8.42$ and $B-V=0.866$
(E. Hog et al. 2000). R. F. Griffin published the orbital solution
of the system in 2015 based on 51 observations, all of which
were determined at the 36-inch observatory at Cambridge
University (R. F. Griffin 2015).

The director of Ramon Maria Aller Observatory, J. A
Docobo, published a practical analytic method to determine
the orbital solution for binary systems. This method calculates
the best orbit of the system that passes through three base
points with a best fit (J. A. Docobo 1985, 2012). In 2014, he
published an important method that allows for the determina-
tion of the orbits of spectroscopic binaries with an astrometric
observation (J. A. Docobo et al. 2014). Furthermore, this
method is considered an effective way to improve the previous
orbital solutions and analyze the consistency of the astrometric
observations. The best orbit is selected based on the better rms
values in $\theta$ and $\rho$ with high coherence with the common orbital
elements from the spectroscopic solution. The new astrometric
orbital elements for HD 130669 are listed in Table 2.

{\it HD 184467, Hip 95995, Gliese 762.1}. The visual observations for 
this system began by Harold A. McAlister at the Kitt
Peak National Observatory (KPNO) in 1983. He used a
speckle interferometry of a 4 m telescope (H. McAlister et al.
1983). In SIMBAD, this system has been classified as a main-sequence 
one with a spectral type of K2 (P. C. Keenan and
R. C. McNeil 1989). In 2010, Farrington used the CHARA
Array to determine the combination of visual-spectroscopic
solutions. He suggested this system to be an SB 2 with a $K2V+K4V$
spectral type for each component with an apparent
magnitude ($mv = 6.59$) and a period of about 494 days
(C. D. Farrington et al. 2010). M Suhail estimated the first
solution using Al-Wardat’s complex method as a close visual-
spectroscopic binary. He presents the individual physical and
geometrical parameters for each system component. The
evolutionary tracks showed the masses to be between 0.8
and 0.9 solar masses for both components as main-sequence
stars with K0V and K1.5V. Isochrones showed this system
with an age of around $9\pm1$ Gyr as low -and intermediate-
mass stars, with a magnitude difference ($\Delta m$) of $0.26\pm0.03$
(S. G. Masda et al. 2016). In 2017, F. Kiefer observed this
system using the spectrograph of the Observatoire de Haute-Provence (SOPHIE) 
and then combined the radial-velocity
measurements with the astrometric measurements. He estimated the masses of 
each component to be $MA=0.833\pm0.031$ and $MB=0.812\pm0.030$, with a parallax of 
$56.10\pm0.81$ mas and with an absolute magnitude of $5.98\pm0.02$ for
the primary component and $6.24\pm0.03$ mag for the secondary
component (F. Kiefer et al. 2018). In 2021, Mitrofanova, using
the radial-velocity amplitudes method, calculated the individual masses 
for HIP 95995 to be $MA=0.99\pm0.11$ and $MB=0.36\pm0.11$ (A. Mitrofanova et al. 2021).

{\it HD 191854, ADS 13461, HIP 99376}. This system has
apparent and absolute magnitudes of $V=7.43$ and $MV=4.3$,
respectively, with $B-V=0.66$, where the difference in
magnitude ($\Delta m$) is 0.74, and a parallax of 0.018 mas
(O. Eggen 1965). In 1969, James W. Christy and R. L.
Walker classified this system as a G5 main-sequence star with
a G4V spectral type for the main component and as a G8V for
the second component, with a difference in magnitude of
$\Delta m=0.5$ (J. W. Christy and R. Walker 1969). W. D. Heintz
gave the absolute magnitudes of the system components to be
8.3 and 9.0 with a spectral type of G4V and an orbital period of
85.61 days, a semimajor axis of $0^{\prime\prime}.449$, an eccentricity of
0.488, an inclination of 116.4, and a dynamical parallax of
0.019 mas (W. Heintz 1997). In 2000, D. Pourbaix described
the spectral type of both system components as G4V and G8V.
The apparent magnitudes of the system were 8.058 and 8.598,
with a mass of 1.2 and 0.9 solar masses. The spectroscopic
orbital parameters are listed in Table 3 ($P=85.2$ , $T=1970.1$,
$a=0.458$, $e=0.492$, $i=115^{\circ}.3$, $\omega=159^{\circ}$, $\Omega=321^{\circ}.7$;
D. Pourbaix 2000).

{\it HD 214222, ADS 16098, HIP 111528}. This is a spectroscopic binary with 
an apparent magnitude of 8.4 and spectral
type of G0V, with a mass of 1.22 solar masses for component
A and 1.16 solar masses for component B. The orbital solution
can be found in Table 3 (D. Pourbaix 2000). G. Starikova
suggested that ADS 16098 has a total mass of 1.9 solar masses,
0.95 for each component. She also suggested an orbital period
of 22.18 yr, a semimajor axis of $0^{\prime\prime}.150$, and a parallax of $0^{\prime\prime}.015$ (G. Starikova 1978). T. Edwards described the Morgan–
Keenan (MK) classification of this system as G1 for both
components with a difference in magnitude of 0.1 and an
apparent magnitude of 8.5 (T. Edwards 1976). E. P. Horch
et al. (2015) determined that the spectral types of the
components are G2V and G3V, with individual masses of
1.00 and 0.97 solar masses, respectively (see Table 3).

\begin{table*}[h]
\centering
\scriptsize
\caption{ Visual orbital elements.}
\label{tab:asthd}
\begin{tabular}{cccccc} \hline\hline
Element & HD 130669 & \multicolumn{1}{l}{HD 184467} & HD 1918154 & HD 214222 \\\hline\hline
P (yr) & 10.01 & 494.307 $\pm$ 0.012 & 85.61 & 22.3455 $\pm$ 0.000 \\
T & 2018.174  & 2013.75  & 1999.85  & 1985.246 $\pm$ 0.000  \\
e  & 0.518  & 0.38933 $\pm$ 0.00029   & 0.488    & 0.362 $\pm$ 0.000    \\
a ($''$) & 0.123  & 0.081 $\pm$ 0.000  & 0.449 & 0.1446 $\pm$ 0.000   \\
i ($^{\circ}$) & 43.5 & 146.15 $\pm$ 0.46  & 116.4  & 63.3 $\pm$ 0.3  \\
$\Omega$ ($^{\circ}$) & 144.8  & 245.72 $\pm$ 0.13  & 142.2 & 110.5 $\pm$ 0.3   \\
$\omega$ ($^{\circ}$) & 335.9 & 180.325 $\pm$ 0.041  & 339.4  & 144.2 $\pm$ 0.0                                \\
Reference & (J. Docobo et al. 2018) & (F. Kiefer et al. 2018) & (W. Heintz1 1997)  & (E. P. Horch et al. 2015)  \\ 
Grade$^{*}$  & 1  & 1  & 2    & 2 \\ 
\hline  \hline
\end{tabular}
\\
	$^*$Sixth Catalog of Orbits of Visual Binary Stars (1: definitive, 5: poor).
\end{table*}

\begin{table*}
\centering
\scriptsize
\caption{Spectroscopic orbital elements.}
\label{tab:sporb11}
\begin{tabular}{cccccc}
\hline	\hline
Element & HD 130669  & HD 184467 & HD 191854  & HD 214222   \\ \hline	\hline
P (d)                                                       & 3619.0 $\pm$ 66.0                                 & 494.313 $\pm$0.012         & 3619.0 $\pm$ 66.0                                                & 8161.68 $\pm$ 40.1766                         \\
P (yr)                                                      & 9.915 $\pm$ 0.005                                 & 1.3534 $\pm$ 0.0001        & 85.215 $\pm$ 0.181                                            & 22.346 $\pm$ 0.110                            \\
T (MJD)                                                     & 54513.0 $\pm$ 11.0                                & 56549.487 $\pm$0.046       & 40476.9 $\pm$ 63.76                                           & 46155.7 $\pm$35.4242                          \\
K$_{1}$ (km/s)                                              & 6.73 $\pm$ 0.07                                   & 9.4911 $\pm$0.0026         & 3.9737 $\pm$ 0.44                                                 & 6.3964$\pm$ 0.14                              \\
K$_{2}$ (km/s)                                              & 6.76 $\pm$ 0.09                                   & 9.733 $\pm$ 0.026          & 5.1183 $\pm$ 0.47                                                 & 6.73626 $\pm$ 0.13                            \\
e                                                           & 0.488 $\pm$ 0.009                                 & 0.38926$\pm$ 0.00030       & 0.4922 $\pm$ 0.0033                                             & 0.36168 $\pm$ 0.013                           \\
$\omega_{1}$ ($^{\circ}$)                                   & 163.0 $\pm$ 2.0                                   & 180.300 $\pm$ 0.043        & 338.93 $\pm$ 1.00                                                    & 144.22 $\pm$ 2.2                              \\
$a_1\sin{i}$ (Gm) & 293 $\pm$ 6 & 59.424 $\pm$0.015 & 1480.46 $\pm$160.00  & 669.279 $\pm$ 15.00 \\
$a_2\sin{i}$ (Gm) & 294 $\pm$ 7  & 60.94$\pm$ 0.16 & 1906.86$\pm$ 180.00 & 704.837$\pm$ 15.00  \\
$a\sin{i}$ (Gm) & 587 $\pm$ 9.2 & 120.36 $\pm$ 0.16 & 3387.32 $\pm$ 240.8 & 1374.112 $\pm$22.10 \\
$\mathcal{M}_{1}$sin$^{3}$i (M$_{\odot}$)                   & 0.308 $\pm$ 0.012                                 & 0.14399 $\pm$ 0.00076      & 0.9034 $\pm$ 0.2000                                            & 0.79917 $\pm$ 0.038                           \\
$\mathcal{M}_{2}$sin$^{3}$i (M$_{\odot}$)                   & 0.306 $\pm$ 0.011                                 & 0.14041 $\pm$0.00038       & 0.7014 $\pm$ 0.1600                                               & 0.75885$\pm$ 0.038                            \\
q                                                           & 1.007 $\pm$ 0.016                                 & 1.0255 $\pm$0.0011         & 1.288 $\pm$ 0.3200                                              & 1.0531 $\pm$ 0.016                            \\
References  & (R. F. Griffin 2012) & (F. Kiefer 2018)& (D. Pourbaix 2000) & (D. Pourbaix 2000) \\

Grade$^{*}$  & 5                                 & 5                                  & 5            & 5 \\ 
 \hline\hline
\end{tabular}
\\
	$^*$ The ninth catalogue of spectroscopic binary orbits (1: poor, 5: definitive).
\end{table*}

\section{Atmospheric and Fundamental Parameters}

In this section, we study the four systems using Al-Wardat’s
method for analyzing binary and multiple systems (M. Al-Wardat
2002b; M. A. Al-Wardat 2012). This computational spectrophotometric 
method uses Kurucz Atlas9 line-blanketed plane-parallel model 
atmospheres of single stars to build the entire SEDs
of the binary and multiple systems. Hence this synthetic
photometry method calculates their synthetic magnitudes and color
indices. The data required to apply this technique are the magnitude
differences obtained either from speckle interferometry measurements or 
from spectroscopic measurements, visual magnitude, and
color indices. These data are widely accessible; therefore, the
technique can be used to study the properties of all binary and
multiple systems, including face-on orbit binaries. This method can
be used even without the system’s orbital information and
spectroscopic data, and that makes it an ideal technique to study
the properties of the systems under investigation. The method was
used to analyze several binary and multiple systems and estimate
their complete set of atmospheric and fundamental parameters of
individual components (e.g., M. Al-Wardat 2007, M. Al-Wardat
and H. Widyan 2009, M. Al-Wardat 2009, M. Al-Wardat 2012,
M. Al-Wardat et al. 2014a, M. Al-Wardat et al. 2014b,
M. A. Al-Wardat et al. 2016, S. G. Masda et al. 2016,
M. Al-Wardat et al. 2017b, S. G. Masda et al. 2018, S. Masda
et al. 2019a, M. A. Al-Wardat et al. 2021a, Y. M. Al-Tawalbeh
et al. 2021, Z. T. Yousef et al. 2021, and M. A. Al-Wardat et al.
2021b).

We now apply the method to each of our systems.

{\it HD 130669, A 2983, HIP 72479}. Having a 0.09 magnitude
difference and using the orbital parallax from this work, Al-
Wardat’s method provides an estimate for the synthetic SEDs
of the entire system and individual components of HD 130669,
as shown in Figure 1. The calculated magnitudes and color
indices of the synthetic SED for the entire system and
individual components of HD 130669 are listed in Table 4.
The estimated atmospheric and fundamental parameters are
listed in Table 5. The effective temperature and luminosity are
used to locate the positions of the components on the
evolutionary tracks of L. Girardi et al. (2000), where the
metallicity used was $Z=0.019$ based on the measured iron abundance 
relative to the Sun $Fe/H=-0.03$ (C. Soubiran et al. 2016). 
The evolutionary tracks of the system using Al-Wardat’s method results 
are shown in Figure 5. The low mass of the system did not allow for 
the estimation of the age of the system.

\begin{figure}
\centering
\includegraphics[width=160mm,height=100mm]{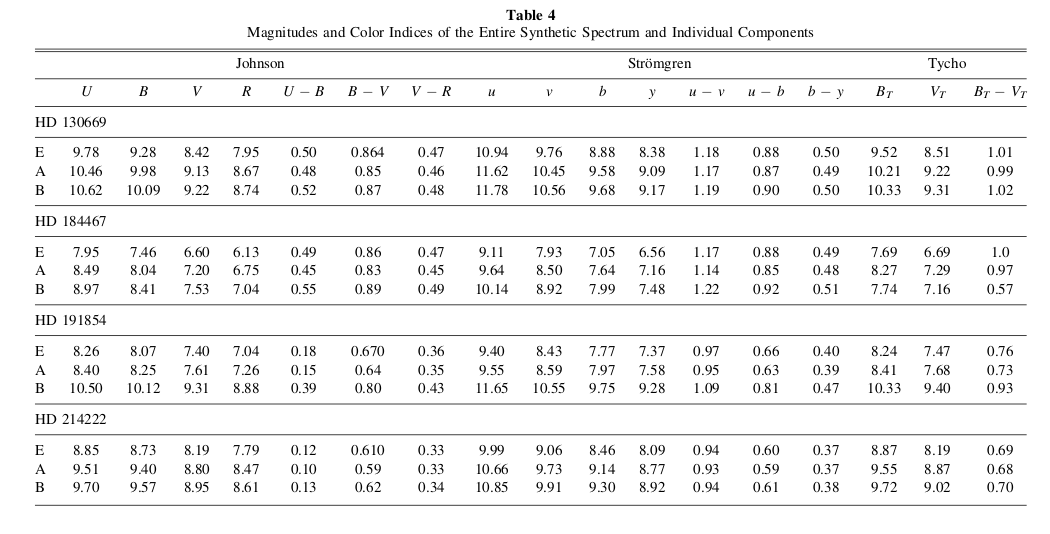}
\end{figure}

\begin{figure}
\centering
\includegraphics[width=90mm,height=70mm]{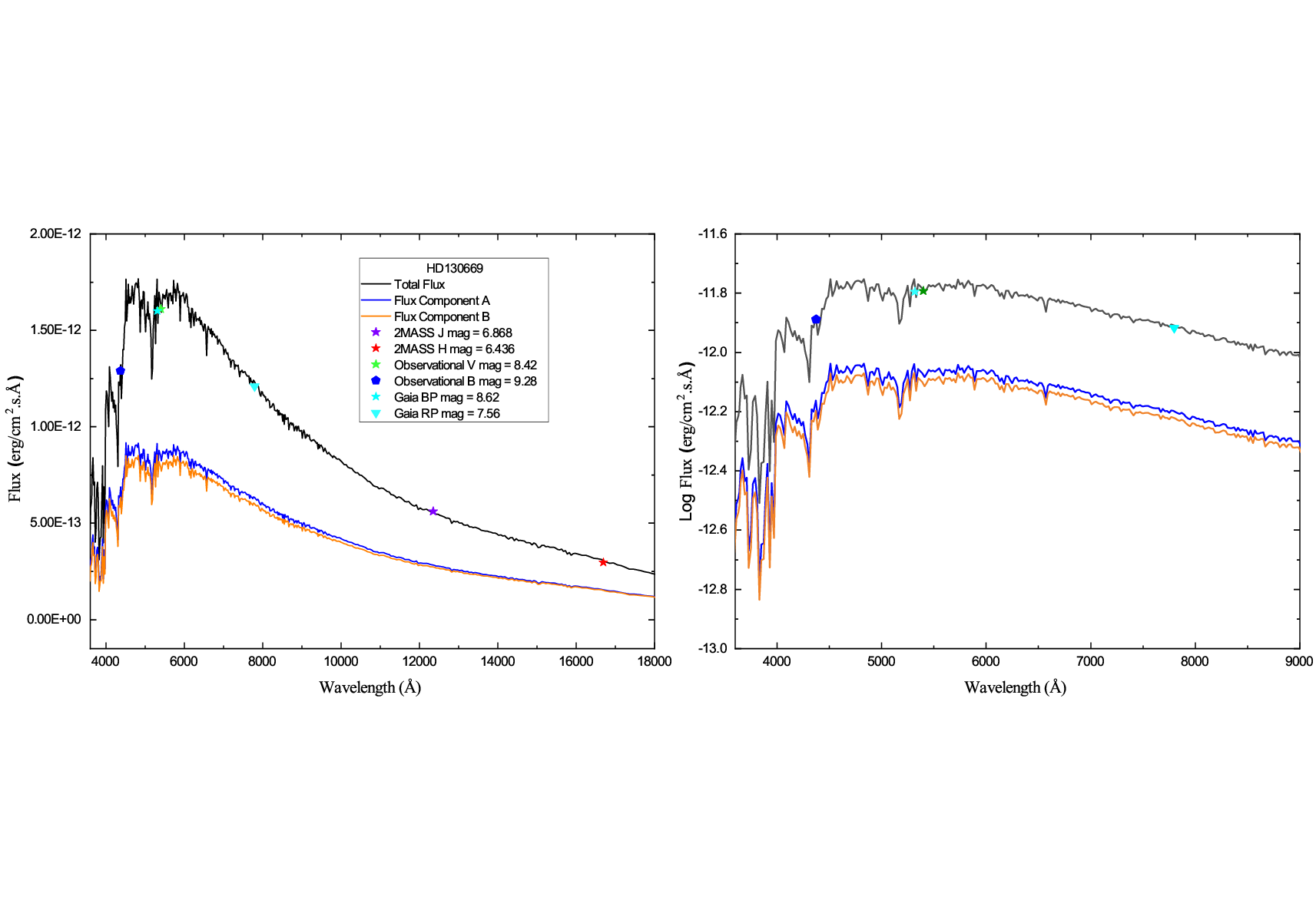}
\caption{Synthetic spectral energy distribution (SED) of the binary system HD 130669 constructed using Al-Wardat’s complex method and Kurucz’s ATLAS9 model atmospheres. The total flux (black), flux of component A (blue), and flux of component B (orange) are shown. Overplotted are observed photometric fluxes from 2MASS (J, H), Gaia (BP, RP), and Johnson photometry (B, V) with their respective magnitudes. This comparison validates the consistency between the synthetic model and observed data across a wide wavelength range ($3,500-18,000 \mathring{A}$).}
\end{figure}

{\it HD 184467, Hip 95995}. Using Al-Wardat’s method, the
synthetic SEDs of the combined and individual components of
HD 184467 are estimated and shown in Figure 2. The
calculated magnitudes and color indices of the synthetic SEDs
of the combined and individual components of HD 184467 are
listed in Table 4. The estimated atmospheric and fundamental
parameters are listed in Table 5. The effective temperature and
luminosity are used to locate the positions of the components
on the evolutionary tracks of L. Girardi et al. (2000), where the
metallicity used was Z = 0.008 based on the measured iron
abundance relative to the Sun $Fe/H=-0.25\pm0.006$
(M. Netopil 2017). Using Al-Wardat’s method results, the
evolutionary tracks of the system are shown in Figure 5. The
best estimate for the age of the system is 6.3 Gyr isochrone for
both components. This suggests that fragmentation is the most
probable formation mechanism of the system.

{\it HD 191854}. Again, using Al-Wardat’s method, we derive
the synthetic SEDs of the combined and individual components of HD 191854. 
These are shown in Figure 3. The
calculated magnitudes and color indices of the synthetic SEDs
of the combined and individual components of HD 191854 are
listed in Table 4, and the estimated atmospheric and
fundamental parameters are listed in Table 5. The effective
temperatures and luminosities of the components were used to
place them on the evolutionary tracks of L. Girardi et al.
(2000), with a metallicity of $Z=0.004$, corresponding to a
measured iron abundance of $Fe/H=-0.05$ (M. Netopil
2017). As shown in Figure 5 and determined using the results
of Al-Wardat’s method, the best-fit isochrone indicates an
estimated system age of approximately 6.3 Gyr for both stars.
This finding supports the likelihood that the system formed
through a fragmentation process.

{\it HD 214222}. As we did with the rest of the systems, we
provide the synthetic SEDs for HD 214222, which can be seen
in Figure 4. The calculated magnitudes and color indices of the
synthetic can be found in Table 4, while the estimated
atmospheric and fundamental parameters are listed in Table 5.
The effective temperature and luminosity of the components
were used to place them on the evolutionary tracks of
L. Girardi et al. (2000), adopting a metallicity of $Z=0.019$,
which corresponds to a measured iron abundance of
$[Fe/H]=-0.05$ (A. Gáspár et al. 2016). As illustrated in
Figure 5 and based on Al-Wardat’s method, the best-fitting
isochrone indicates an age of approximately 3.548 Gyr for
both components. This result suggests that the system most
likely formed through the fragmentation mechanism.

\begin{figure}
\centering
\includegraphics[width=90mm,height=70mm]{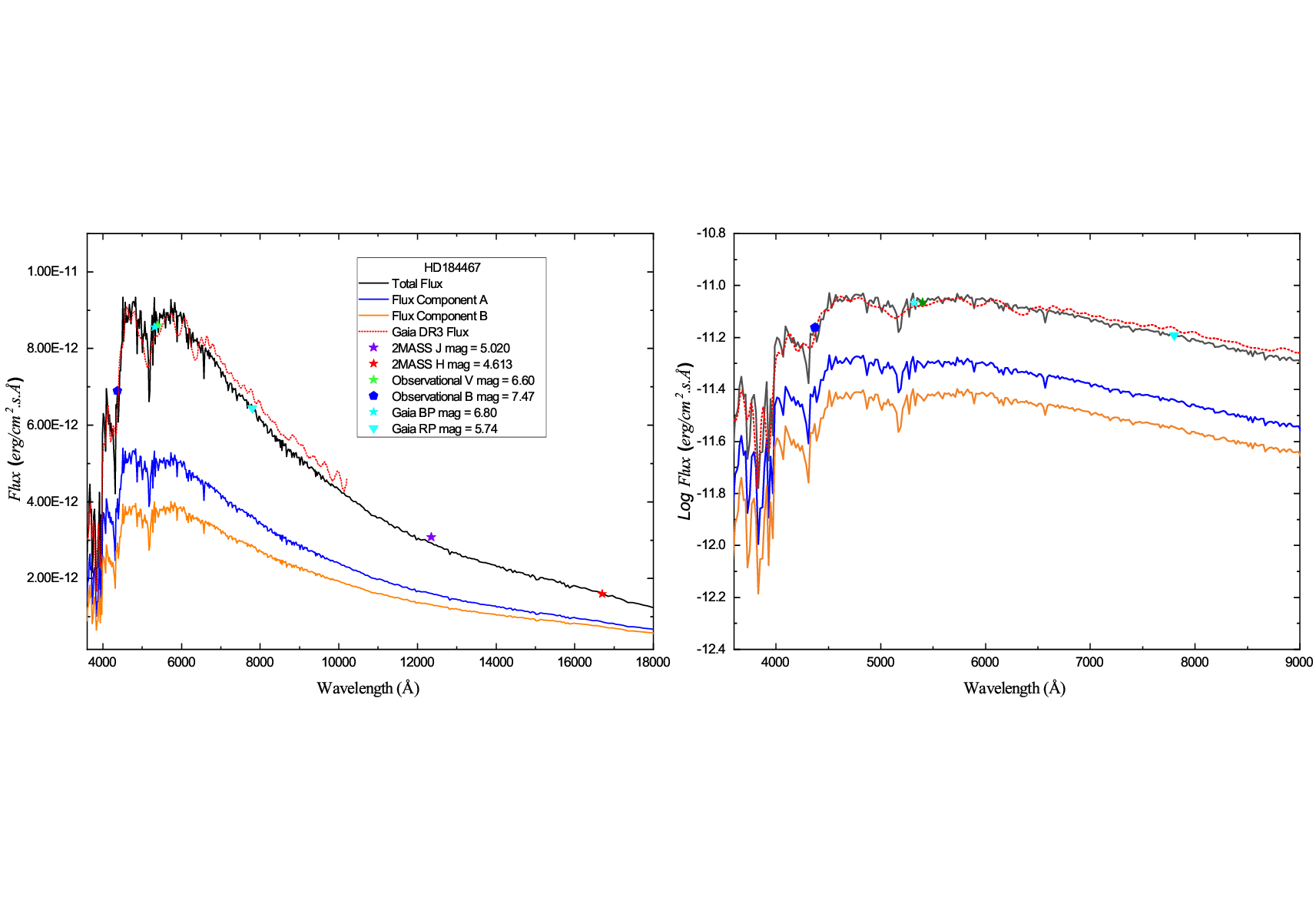}
\caption{Synthetic spectral energy distribution (SED) of the binary system HD 184467 constructed using Al-Wardat’s complex method and Kurucz’s ATLAS9 model atmospheres. The total flux (black), flux of component A (blue), and flux of component B (orange) are shown. Overplotted are observed photometric fluxes from 2MASS (J, H), Gaia (BP, RP), and Johnson photometry (B, V) with their respective magnitudes. This comparison validates the consistency between the synthetic model and observed data across a wide wavelength range ($3,500-18,000 \mathring{A}$).}
\end{figure}

\begin{figure}
\centering
\includegraphics[width=90mm,height=70mm]{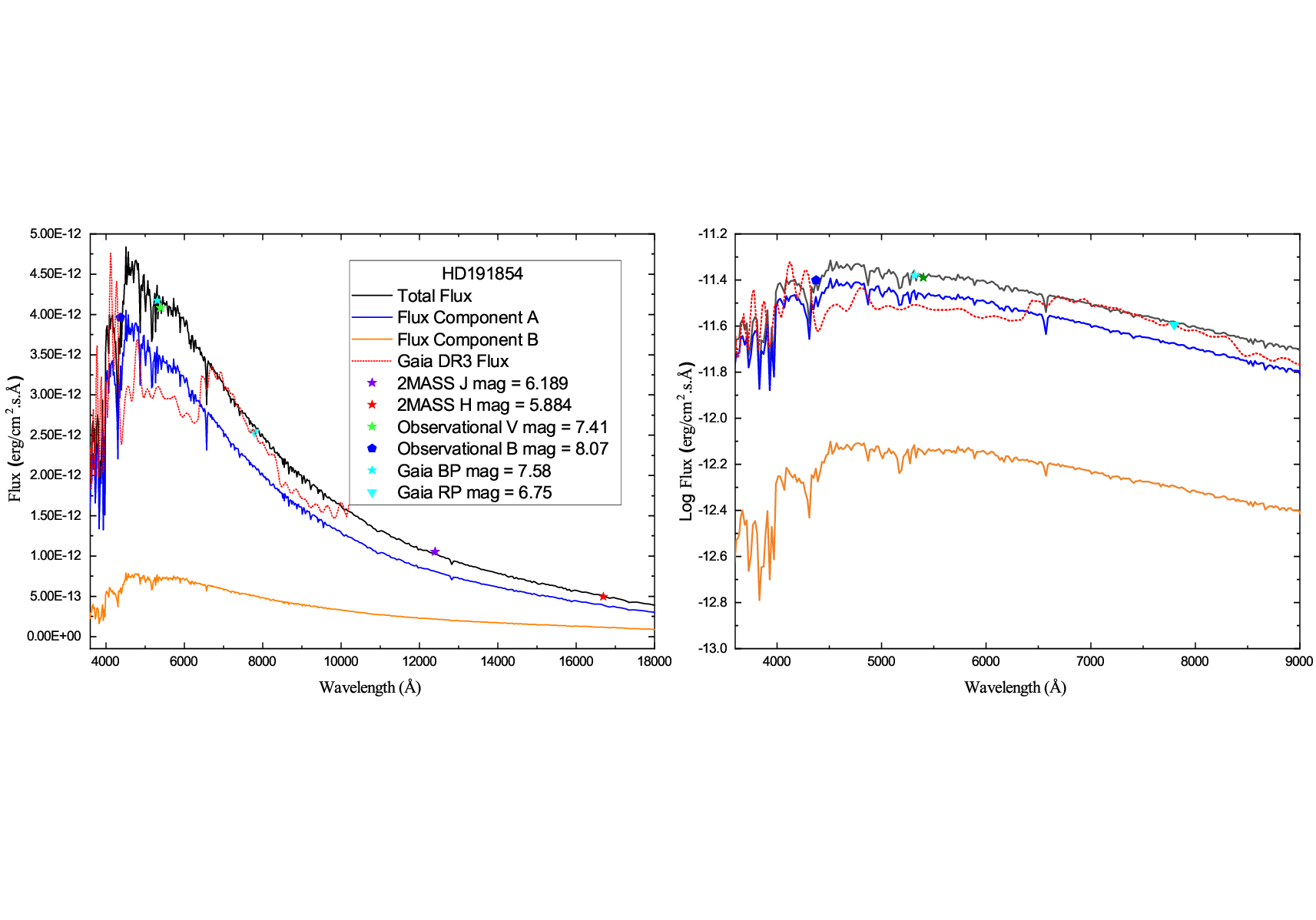}
\caption{Synthetic spectral energy distribution (SED) of the binary system HD 191854 constructed using Al-Wardat’s complex method and Kurucz’s ATLAS9 model atmospheres. The total flux (black), flux of component A (blue), and flux of component B (orange) are shown. Overplotted are observed photometric fluxes from 2MASS (J, H), Gaia (BP, RP), and Johnson photometry (B, V) with their respective magnitudes. This comparison validates the consistency between the synthetic model and observed data across a wide wavelength range ($3,500-18,000 \mathring{A}$).}
\end{figure}

\begin{figure}
\centering
\includegraphics[width=90mm,height=70mm]{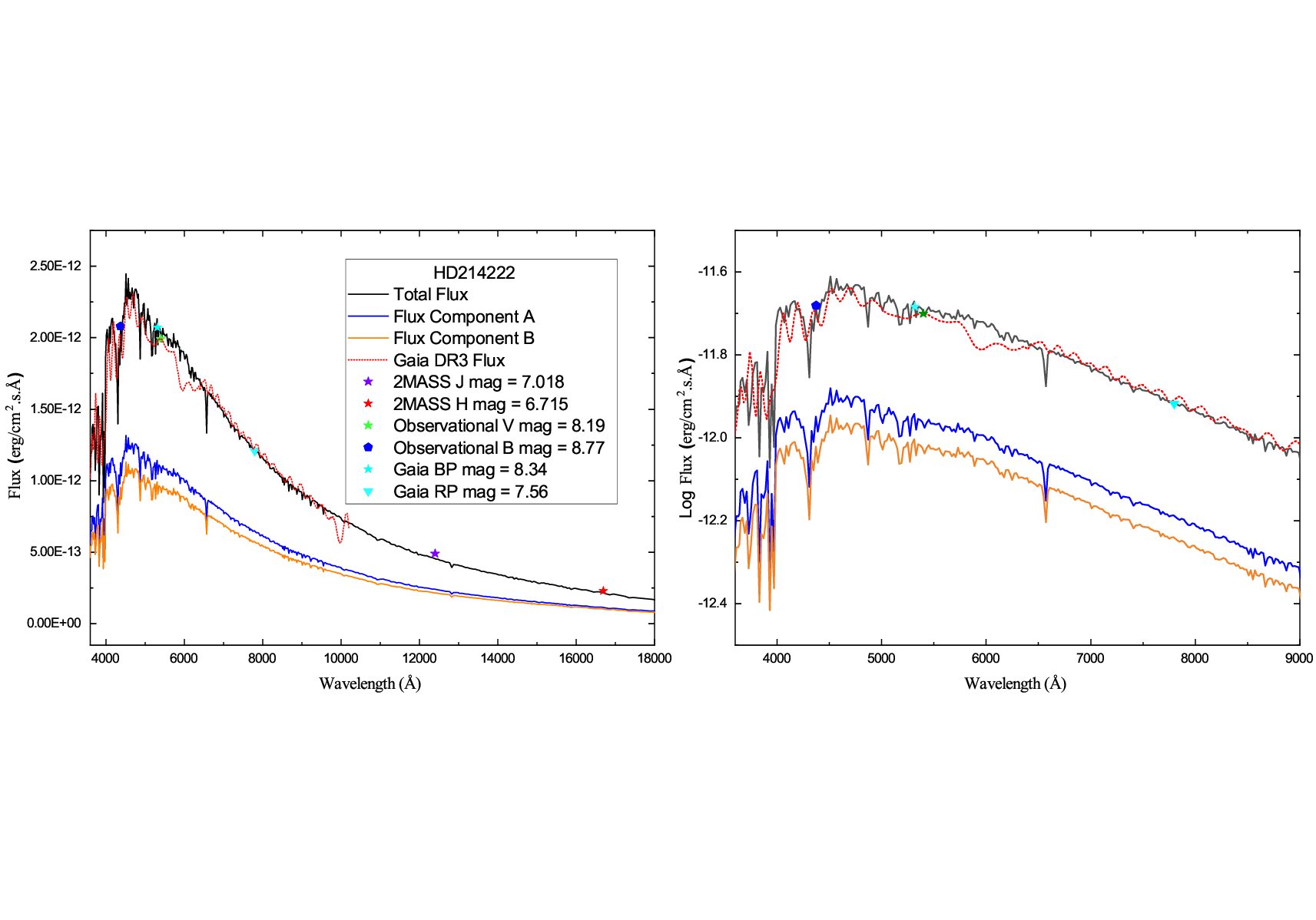}
\caption{Synthetic spectral energy distribution (SED) of the binary system HD 21422 constructed using Al-Wardat’s complex method and Kurucz’s ATLAS9 model atmospheres. The total flux (black), flux of component A (blue), and flux of component B (orange) are shown. Overplotted are observed photometric fluxes from 2MASS (J, H), Gaia (BP, RP), and Johnson photometry (B, V) with their respective magnitudes. This comparison validates the consistency between the synthetic model and observed data across a wide wavelength range ($3,500-18,000 \mathring{A}$).}
\end{figure}

\begin{figure}
\centering
\includegraphics[width=70mm,height=60mm]{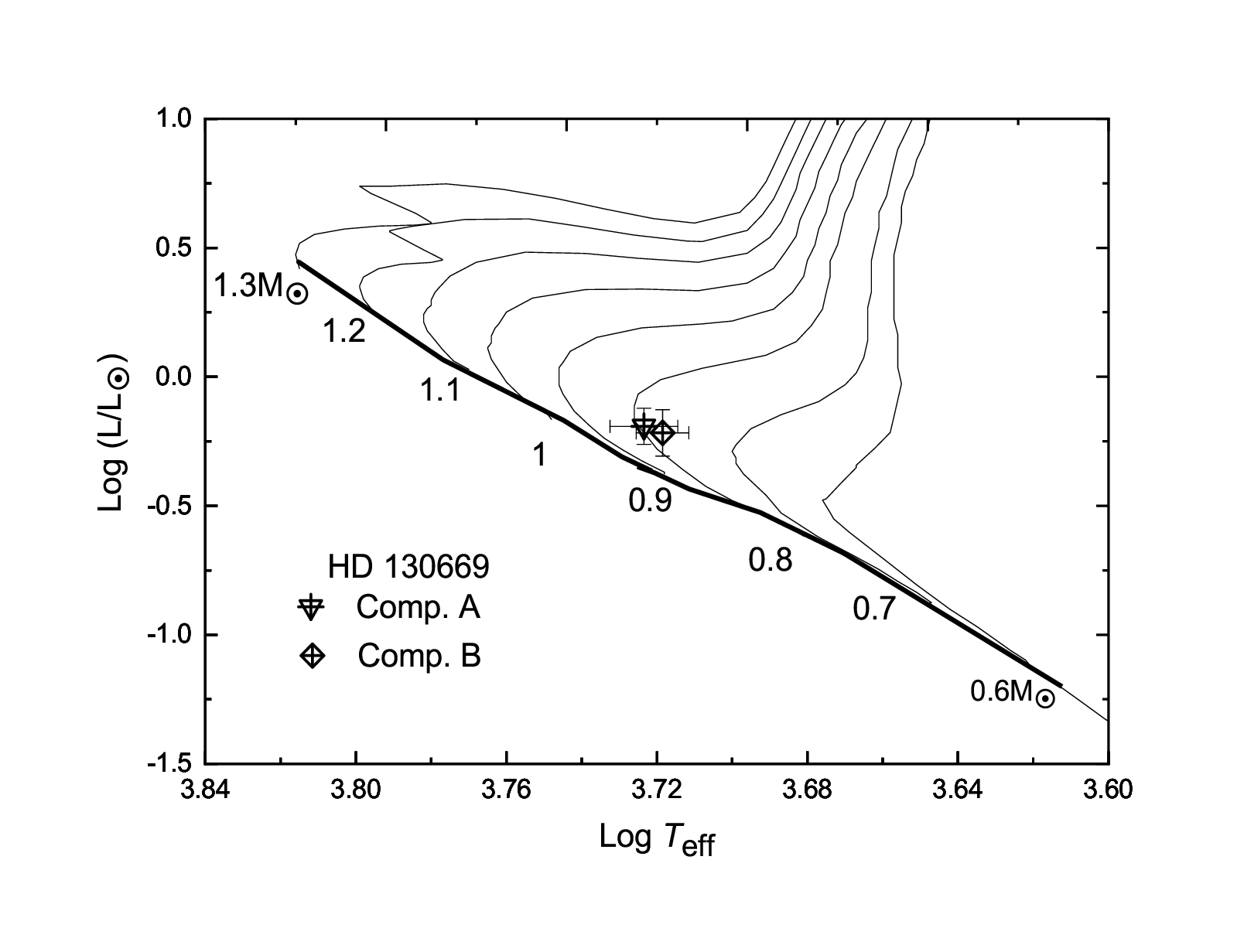}
\includegraphics[width=70mm,height=60mm]{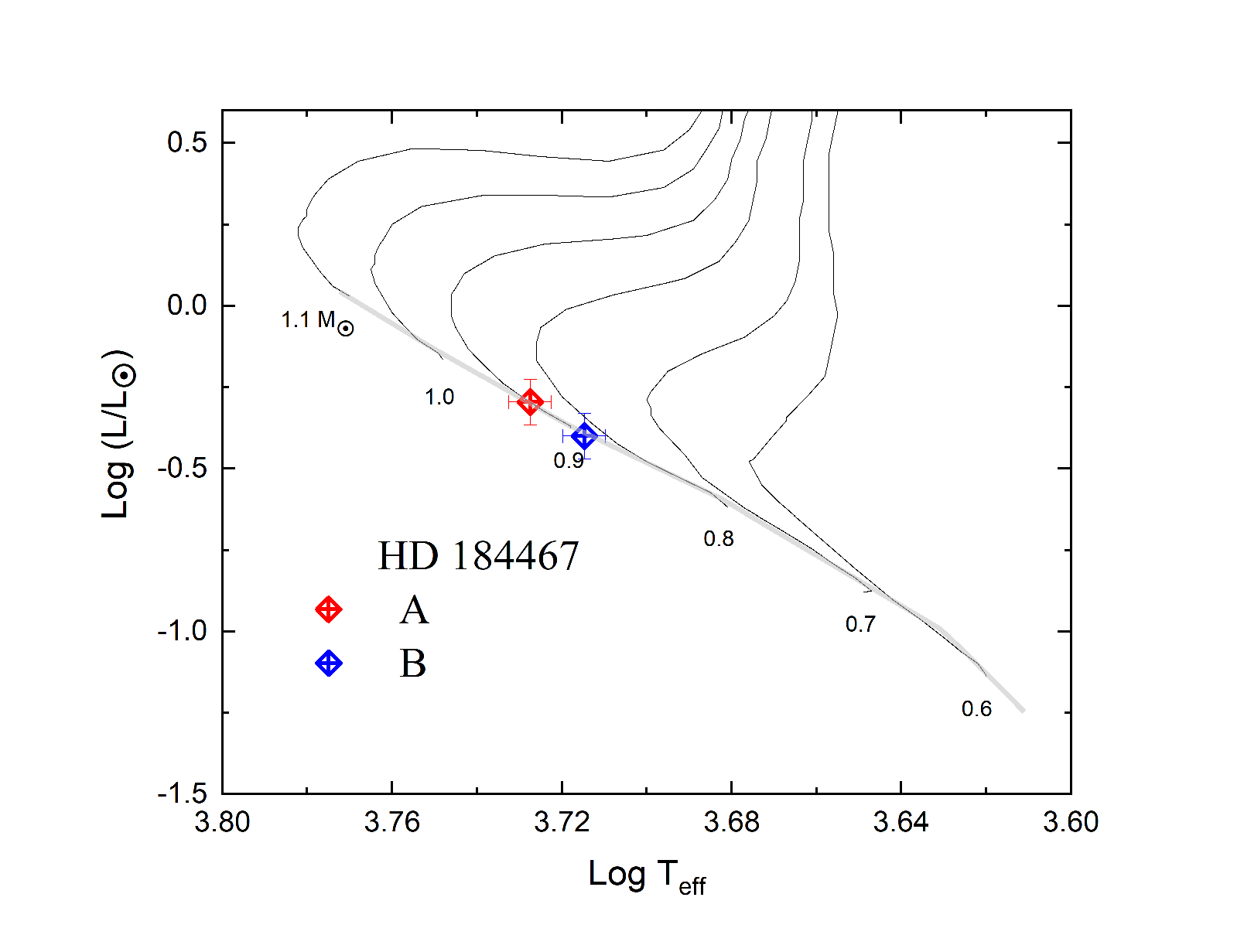}
\includegraphics[width=70mm,height=60mm]{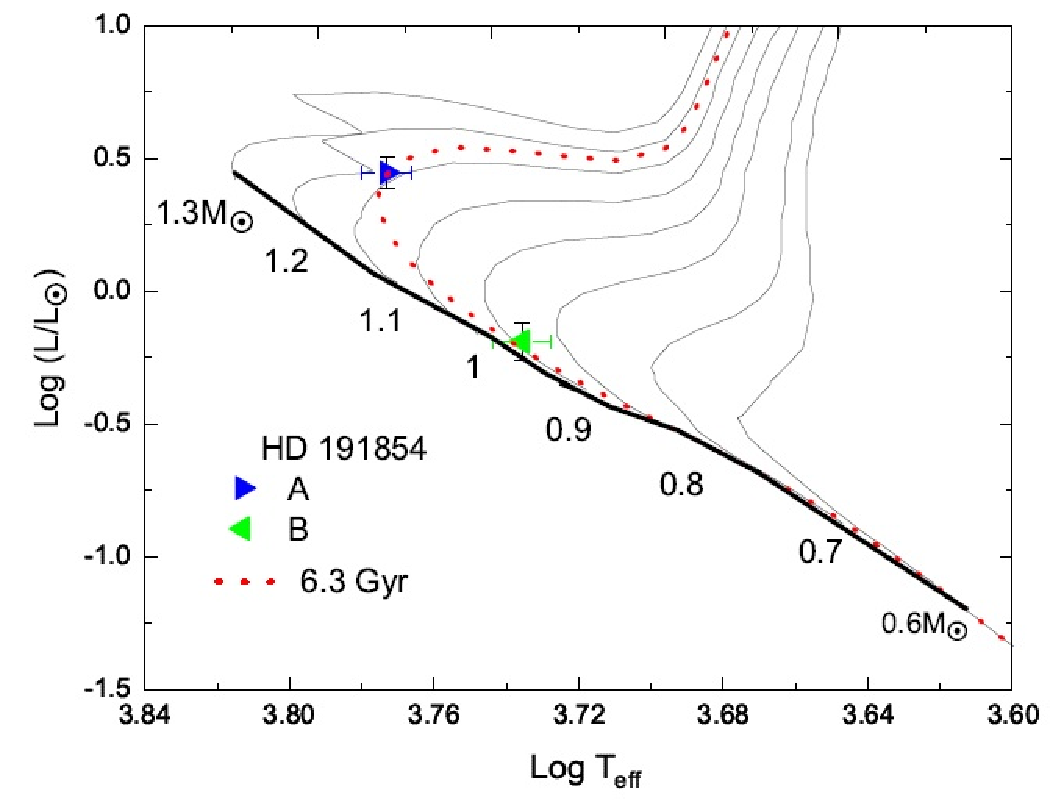}
\includegraphics[width=70mm,height=60mm]{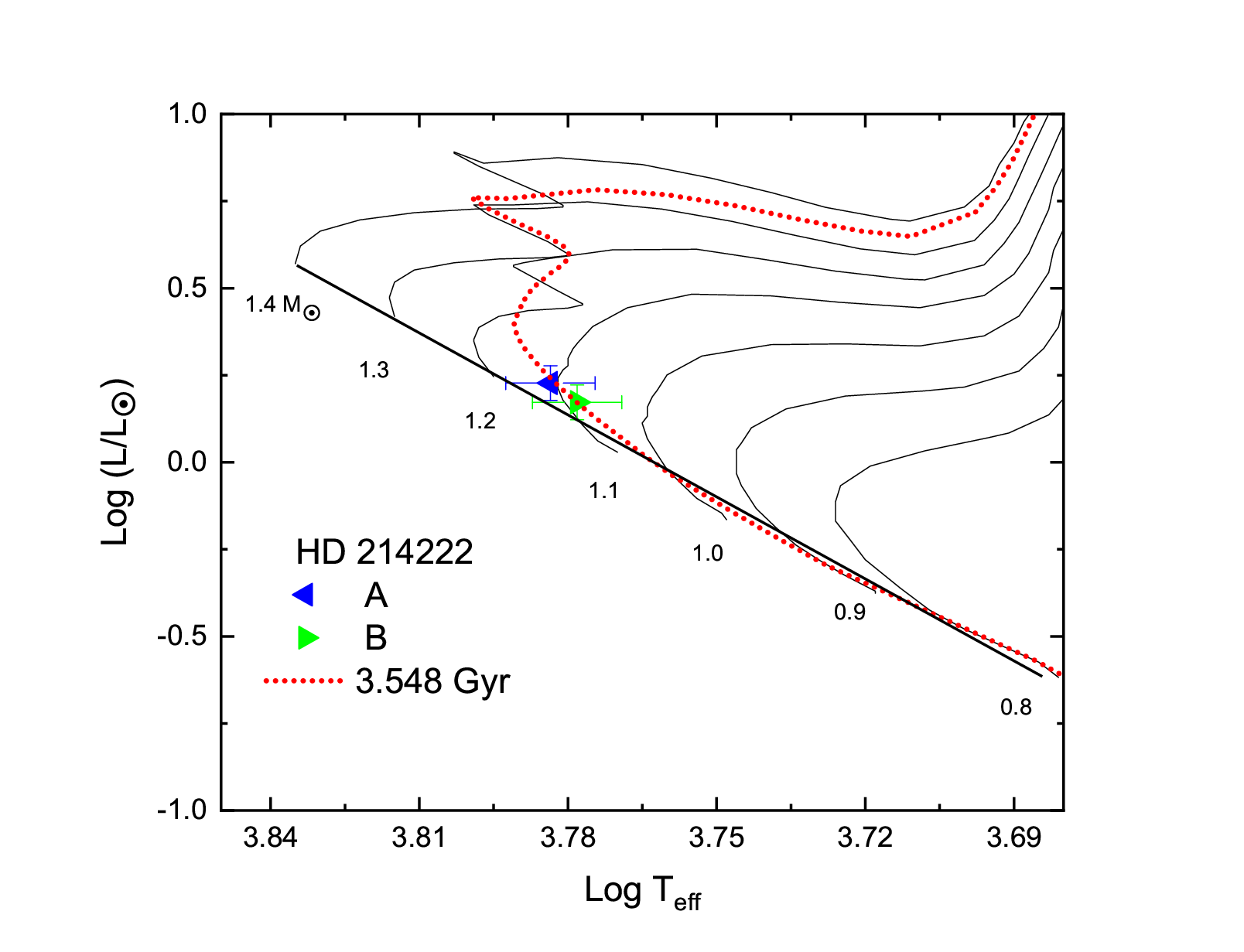}   
\caption{ H–R diagrams for the binary systems HD 130669, HD 184467, HD 191854, and HD 214222. The positions of the individual stellar components (A and B) are plotted with their respective uncertainties. Evolutionary tracks corresponding to different stellar masses are shown by black lines, while the system's best-fitting isochrone (age) is indicated by a red dotted line. The Zero-Age Main Sequence (ZAMS) is depicted as a thick solid black line. These diagrams were constructed using the evolutionary models to determine the masses and ages of the components based on their effective temperatures and luminosities.}
\end{figure}

\begin{figure}
\centering
\includegraphics[width=140mm,height=80mm]{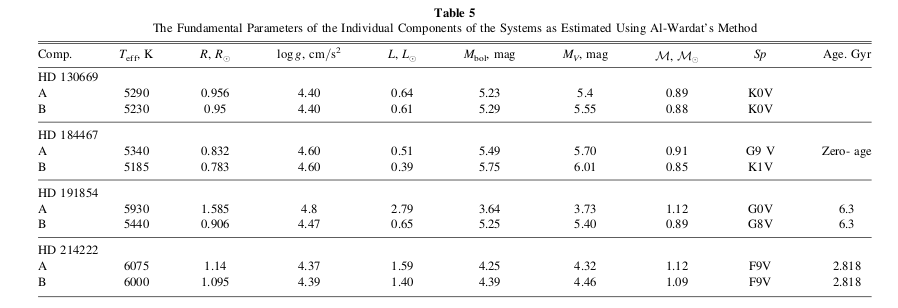}
\end{figure}

\section{Parallax and Masses}

By studying double-lined spectroscopic binaries, we can
accurately determine stellar masses and orbital parameters. We
can calculate the entire orbital geometry by combining
spectroscopic radial-velocity measurements with visual or
interferometric orbit measurements, including the inclination
and mass ratio. Thus absolute masses can be determined
without the ambiguity of spectroscopic data alone. By
combining angular and linear orbital measurements, orbital
parallax can be calculated, offering a direct and reliable
method to determine stellar distances. This approach also helps
validate and refine astrometric catalogs like Hipparcos and
Gaia. Then, by combining synthetic photometry with modeling
of how energy is distributed across different wavelengths, this
method allows for a detailed understanding of the two stars in
a binary system. This understanding is from both a physical
and atmospheric perspective. This means that using spectroscopic and visual methods together is a strong and accurate
way to study SB2 systems. This can help us improve models of
how stars evolve and advance our understanding of stellar
astrophysics.

We can calculate the total mass of the system using Kepler's third law:  
\begin{equation}
    \mathcal{M}_{1}+\mathcal{M}_{2}=\left ( \frac{a''}{\pi ''} \right )^{3}\frac{1}{P^{2}},
\end{equation}
where $\mathcal{M}_1$ and $\mathcal{M}_1$ are the masses of the two stars, $a$ is the semimajor axis in arcseconds, P is the period in years, and $\pi$ is the Gaia parallax. Also, we can obtain the mass ratio for the double-lined spectroscopic binaries from the orbital solution using

\begin{equation}
    q=\frac{\mathcal{M}_{1} sin^{3}(i)}{\mathcal{M}_{2} sin^{3}(i)}= \frac{\mathcal{M}_{1}}{\mathcal{M}_{2}}
\end{equation}
where $i$ is the inclination in degrees. Then, by using equations (1) and (2), we can determine the most precise individual mass of each component:
\begin{equation}
  \mathcal{M}_{1} =  q * \mathcal{M}_{2}, \end{equation}
\begin{equation}
  \mathcal{M}_{2}=\left ( \frac{a''}{\pi ''} \right )^{3}\left ( \frac{1}{P} \right )^{2}\left ( \frac{1}{1+q} \right ).
\end{equation}
 Considering $A=a\sin{i}$, $A_{1}=a_{1}\sin{i}$,  $A_{2}=a_{2}\sin{i}$
\begin{equation}
  A_{a.u}= \frac{(A_{1} + A_{2})_{km}}{149 597 870.700}. 
  \label{eq:orbp}
\end{equation}
Stellar masses are influenced significantly by uncertainties in orbital inclination (i) and orbital parallax ($\pi$). Based on the angular semi-major axis and parallax, the physical semi-major axis can be calculated as follows:
\begin{equation}
    \pi=\frac{a''}{a}.
\end{equation}

By adding the $a^3$ term to Kepler’s third law, even small
parallax errors are magnified, resulting in greater mass
uncertainty. Furthermore, Kepler’s third law includes a triple
dependence term, which magnifies even small errors in
parallax, thus increasing mass uncertainty. Nonlinear errors
can also occur due to deviations in inclination, especially when
orbits are not edge-on. As shown in Table 6, we propagated
these uncertainties using standard partial derivative-based
error propagation methods. Therefore, deviations in inclination, 
especially for orbits that are not edge-on, can cause
nonlinear errors in mass estimations. We apply the previous
equations to determine the individual masses and orbital
parallax of the double-lined spectroscopic binary HD 184467.

{\it HD 184467, ADS 13529, HIP 96295}. We apply the same
combined methodology of visual and spectroscopic analysis to
the system HD 184467 in order to determine its parallax and
individual stellar masses. Using the orbital solution and
inclination angle from the visual orbit ($i=146^{\circ}.15\pm1^{\circ}.5$),
and calculating $a\sin{i}$ from the spectroscopic orbit, we find the
semimajor axis of the system to be a = 1.444 ± 0.0042 au.

The visual orbit provides an angular semimajor axis of $a^{\prime\prime}= 0^{\prime\prime}.081 \pm 0^{\prime\prime}.001$. From this, we calculate the orbital parallax using 
\begin{displaymath}
\pi_{\text{orb}} = \frac{a''}{a} = \frac{0.081}{1.444} = 0''.05610 \pm 0''.00071,
\end{displaymath}
which corresponds to a distance of approximately 17.82 pc. This is in excellent agreement with the orbital parallax provided in the orbital solution and confirms the precision of the combined visual-spectroscopic model.

Using Kepler’s third law with this parallax and the orbital period $P = 1.3537~\text{years}$, we determine the total mass of the system to be  
\[
\mathcal{M}_{\text{total}} = \left( \frac{a''}{\pi''} \right)^3 \cdot \frac{1}{P^2} = \left( \frac{0.081}{0.05610} \right)^3 \cdot \frac{1}{(1.3537)^2} = 1.6451~\mathcal{M}_{\odot}.
\]

From the spectroscopic mass ratio $q = \mathcal{M}_1 / \mathcal{M}_2 = 1.0254$, we derive the individual component masses:  
\[
\mathcal{M}_2 = \frac{\mathcal{M}_{\text{total}}}{1 + q} = \frac{1.6451}{1 + 1.0254} = 0.8122 \pm 0.0107~\mathcal{M}_{\odot},
\]
\[
\mathcal{M}_1 = q \cdot \mathcal{M}_2 = 1.0254 \times 0.8122 = 0.8329 \pm 0.0188~\mathcal{M}_{\odot}.
\]

These results are in strong agreement with Al-Wardat’s method, which yields  
\[
\mathcal{M}_1 = 0.91 \pm 0.05~\mathcal{M}_{\odot}, \quad \mathcal{M}_2 = 0.85 \pm 0.04~\mathcal{M}_{\odot}.
\]
 This consistency between orbital dynamics and atmosphere modeling confirms the reliability of the adopted method.

The dynamical parallax based on Al-Wardat’s masses and the orbital elements is:  
\[
\pi_{\text{dyn}} = 0''.05630 \pm 0''.0016,
\]
 The close match between the orbital, dynamical, and catalogue parallaxes demonstrates the strength of the combined approach for accurately characterizing SB2 systems.

{\it HD 130669, A 2983, HIP 72479:}
The $V$ magnitude of HD 130669, as provided by SIMBAD, is 8.42, with $B-V=0.866$ (M. Perryman et al. 1997; F. van Leeuwen 2007).
 We apply the same methodology for HD 130669 and we estimate the difference in magnitudes to be 0.5, which agrees with Griffin's suggestion (M. Al-Wardat 2012; R. F. Griffin 2015). 
  The total absolute magnitude equals 5.19 $\pm$ 0.04, while the absolute magnitudes for the components are 5.00$\pm$ 0.02 and 5.50 $\pm$ 0.02, respectively. R. F. Griffin (2015) suggests K0V spectral types for both components with 2-3 subtypes in the accuracy.
The total mass is 1.92 $\pm$ 0.09, which is close to 1.85 $\pm$ 0.08 (of this work). The mass ratio is 1.03 $\pm$ 0.034, which is in good agreement with Griffin's calculation of 1.005 $\pm$ 0.016 (R. F. Griffin 2015). The inclination is equal to $42.48^{o}$ $\pm$ $2.1^{o}$, which is very close to $43.5^{o}$ $\pm$ $0^{o}$ of this work.

M. Al-Wardat (2012) analyzed the system using the Hipparcos 1 parallax and estimated the masses of the individual components to be:
\begin{equation}
\label{eq:masses}
\mathcal{M}_{1} = 0.94 \pm 0.05 \mathcal{M}_{\odot} \hspace{0.1cm} \mbox{and} \hspace{0.1cm} 
\mathcal{M}_{2} = 0.85 \pm 0.04 \mathcal{M}_{\odot}.
\end{equation}
M. Al-Wardat (2012) used a 0.5 magnitude difference from speckle interferometry. The speckle magnitude difference gives a mass ratio of about 1.10, different from the spectroscopic mass ratio of 0.9 in Esa (1997), which gives a 0.2 difference in the mass ratio.

Here, we take the same steps to determine 
$a\sin{i}$ in au, and the visual orbit inclination $i=43.5^{\circ} \pm 0^{\circ}$. This yields a semimajor axis $a = 5.7003 \pm 0.0615$ au.
Combining this result with the semimajor axis of $0.123 \pm 0.000$ arcseconds from the visual solution, we can then deduce the orbital parallax of the system to be $\pi_{orb} =0.0216'' \pm 0.0015''$, which agrees with the Hipparcos parallax of $0.02259'' \pm 0.00125''$ (F. van Leeuwen 2007).
Using this new parallax, the total mass is $\mathcal{M}_{total} = 1.8486 \pm 0.0815$ $\mathcal{M}_{\odot}$. Hence, using the spectroscopic solution, the mass ratio and the individual masses of the system are $q = 1.005 \pm 0.016$,
$\mathcal{M}_{1} = 0.9266 \pm 0.0511 \mathcal{M}_{\odot} $  and
$\mathcal{M}_{2} = 0.9220 \pm 0.0448 \mathcal{M}_{\odot}$ respectively. On the other hand, using Al-Wardat's method,
we find for the total mass and the individual masses of the system $q=1.01 \pm  0.016$,
$\mathcal{M}_{1} = 0.89 \pm 0.05 \mathcal{M}_{\odot}$ and
$\mathcal{M}_{2} = 0.88 \pm 0.04 \mathcal{M}_{\odot}$ respectively.
Both solutions agree with each other within the error ranges. 
The calculated dynamical parallax of HD 130669 using Al-Wardat's masses and the orbital parameters of this work is $0.02180'' \pm 0.0025''$, i.e the system is at a distance of $45.85pc$. On the other hand, the orbital parallax puts the system at a distance of $46.296 pc$. This assures the coincidence between the two solutions.

{\it HD 191854, ADS 13461, HIP 99376:}
In this paragraph, we follow the same procedure as previously in order to determine the masses and parallax of this system.  
The procedure gives us a semimajor axis $a=25.2768 \pm 1.7911$ au, while from the visual solution we obtain $a=0.449 \pm 0.0001$ arcseconds.
Based on that, we deduce the orbital parallax of the system $\pi_{orb} = 0.0178'' \pm 0.0013''$, which agrees with the Gaia DR3 parallax of $0.01929'' \pm 0.00128''$ (C. Fontanive and D. Bardalez Gagliuffi 2021).
Using this new parallax, the total mass of the system is 
$\mathcal{M}_{total}$=2.224$\pm$ 0.0309$\mathcal{M}_{\odot}$.
From the spectroscopic solution, the mass ratio and the individual masses of the system  are 
$q = 1.268 \pm  0.014$, $\mathcal{M}_{1} = 1.2438 \pm 0.1377 \mathcal{M}_{\odot}$ and $\mathcal{M}_{2} = 0.9802\pm 0.0276 \mathcal{M}_{\odot}$ respectively.  At the same time, we obtain from Al-Wardat's method $q=1.25 \pm 0.022$, $\mathcal{M}_{1} = 1.12 \pm 0.06 \mathcal{M}_{\odot}$ and $\mathcal{M}_{2} = 0.89 \pm 0.04 \mathcal{M}_{\odot}$.
Both solutions agree with each other within the error ranges. 

{\it HD 214222, ADS 16098, HIP 111528:}
Following the same steps as for the other systems,
the binary semimajor axis is $10.2810 \pm 0.1668$ au, while from the visual solution we get $a=0.1446 \pm 0.0001$ arcseconds. Hence, we calculate the orbital parallax of the system which is found to be $\pi_{orb} = 0.01410'' \pm 0.0023''$ and is in agreement with the Gaia DR2 parallax of $0.01435'' \pm 0.00319''$ (A. G. Brown et al. 2016).
Using this parallax value, the combined mass of the system is
$\mathcal{M}_{t}$=1.1165$\pm$ 0.1958$\mathcal{M}_{\odot}$.
Finally, using the spectroscopic solution, the mass ratio and the individual masses of the system are $q = 1.044 \pm  0.022$, $\mathcal{M}_{1} = 1.1165 \pm 0.1958 \mathcal{M}_{\odot}$ and $\mathcal{M}_{2} = 1.0662 \pm 0.1098 \mathcal{M}_{\odot}$.
The Al-Wardat's method for this systems gives  $q =  1.01 \pm 0.016$, $\mathcal{M}_{1} = 1.12 \pm 0.06 \mathcal{M}_{\odot}$ and $\mathcal{M}_{2} = 1.09\pm 0.05 \mathcal{M}_{\odot}$.  Both solutions agree with each other within the error ranges. 

\subsection{Comparison of Mass Ratios}

Most systems show strong consistency in mass ratios when comparing calculated and observed values. The results confirm the reliability of the methods used to calculate the masses of binary stars. The mass ratios ($q_{calc} $) obtained from the combination of astrometric and spectroscopic orbital elements closely match those obtained from solely spectroscopic measurements of radial velocity. For HD 130669, the values are nearly identical ($q_{calc} $ = 1.005 $\pm$ 0.034; $q_{spec} $= 1.007 $\pm$ 0.016), as is the case of HD 184467 ($q_{calc} $ = 1.026 $\pm$ 0.022; $q_{spec} $= 1.0255 $\pm$ 0.0011) and HD 214222 ($q_{calc} $ = 1.047 $\pm$ 0.049; $q_{spec} $= 1.0531 $\pm$ 0.016); see Table 7 (and Table 6), with all differences well within the respective uncertainties. Consequently, the physical parameters derived appear to be highly credible. The discrepancy between the calculated and spectroscopic mass ratio ($q_{calc} $ = 1.268$\pm$0.058; $q_{spec} $= 1.288$\pm$0.320) for HD 191854 is the largest among the four systems analyzed. Despite the statistical consistency within uncertainties 1$\sigma$, the relatively large error in the spectroscopic ratio suggests potential challenges for the radial velocity solution. There are several explanations for this, including blended spectral lines, moderate orbital inclination reducing velocity amplitudes, and a lack of resolution in the data. Thus, apparent discrepancies are more a reflection of observational uncertainty than methodological inconsistencies. It is recommended to acquire higher resolution spectra and extend temporal coverage to improve the precision of the system's spectroscopic parameters.

\begin{figure}
\centering
\includegraphics[width=150mm,height=70mm]{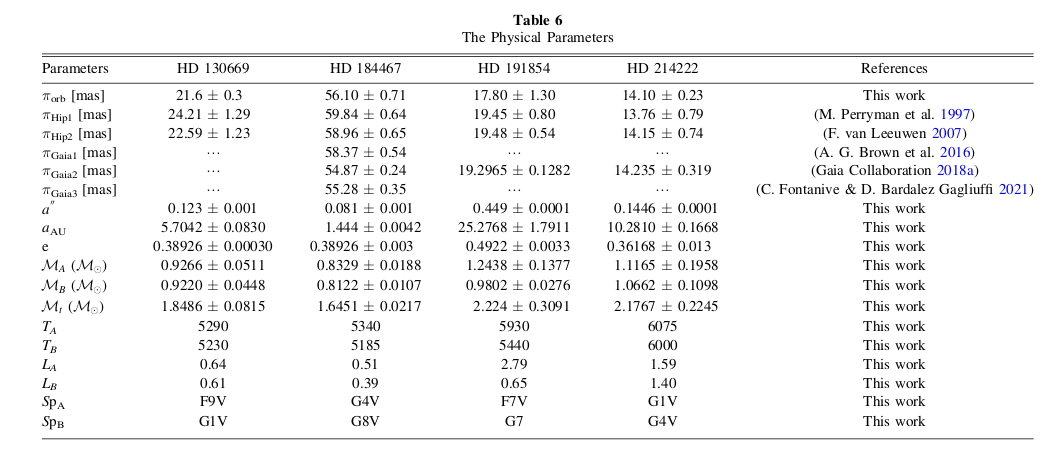}
\end{figure}

\begin{figure}
\centering
\includegraphics[width=110mm,height=40mm]{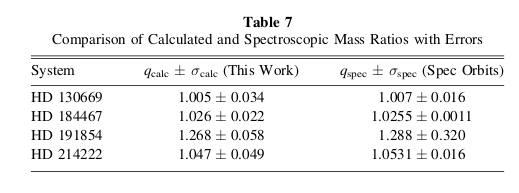}
\end{figure}

\section{Dynamical Stability and Habitability}
Since we have determined the masses and orbital parameters
of our binary systems in the previous sections, here we
investigate whether these systems may provide a suitable
dynamical environment for hosting planets and whether any of
these planets can reside in the habitable zones of the respective
systems.

\subsection{Dynamical Stability}
Although some of the systems may favor the existence of
circumbinary planets (the planet being on an orbit around both
stars) while some other may favor circumstellar orbits (the
planet orbiting one of the stars), we are going to explore both
configurations for each system. For the purpose of searching
for stable circumbinary orbits, we make use of the work of
N. Georgakarakos et al. (2024), while for circumstellar orbits
we use the respective formula provided by M. J. Holman and
P. A. Wiegert (1999).

For circumbinary planetary orbits, we evaluate the formula
for the upper critical semimajor axis (the semimajor axis
beyond which all planetary orbits are stable) from N. Georgakarakos et al. (2024). For orbits of planetary eccentricities 0.8
or less, the formula is given by
\begin{equation}
a^{cr}_{o}= a \,\cdot 10^k,
\end{equation}
where a is the binary semi-major axis. For coplanar orbits, the exponent $k$ is given by
\begin{eqnarray}
k &=& 0.23612-0.29377\log_{10}(M_b)+1.06753e+0.62109e_p-0.21512[log_{10}(M_b)]^2-1.52936e^2-\nonumber\\
& & -0.47480e_p^2-0.03932[\log_{10}(M_b)]^3+0.87506e^3+1.25895e_p^3,
\end{eqnarray}
with $M_b=\mathcal{M_B}/(\mathcal{M_A}+\mathcal{M_B})$ and $e$ and $e_p$ are the eccentricities of the binary and the planet, respectively.  For initially circular planetary orbits, i.e. $e_p=0$, the upper critical semi-major axes can be found in Table 8.

\begin{figure}
\centering
\includegraphics[width=110mm,height=40mm]{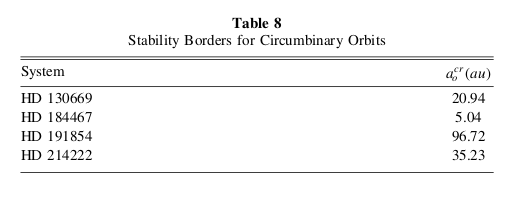}
\end{figure}

In the event that the planetary orbit is assumed to be initially
eccentric, then the stability border will be farther away from
the stellar binary. This can be seen in Figure 6.

\begin{figure}
\begin{center}
\includegraphics[width=88mm]{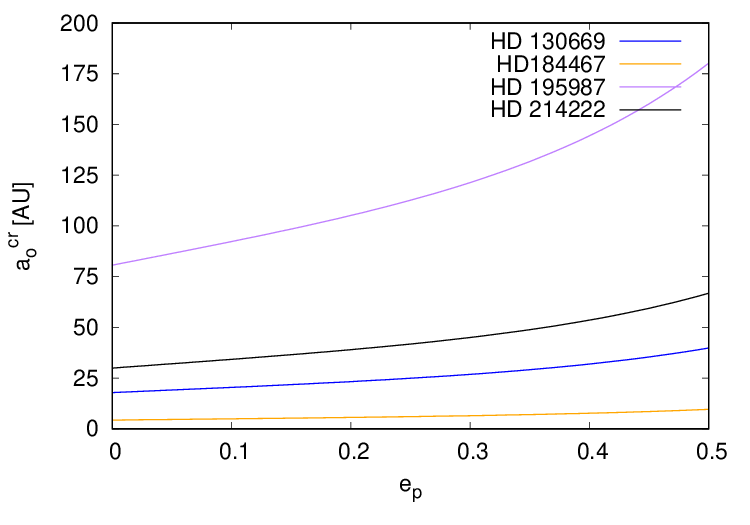}
\caption{Circumbinary stability curves for the four binary systems HD 130669, HD 184467, HD 191854, and HD 214222. The curves show the critical semi-major axis a$_{cr}$ of a stable circumbinary planet as a function of planet eccentricity e$_{p}$, calculated based on the orbital and stellar parameters of each system.}
\label{figcbs}
\end{center}
\end{figure}

In order to determine the stability status of circumstellar
orbits in our binary systems, we make use of the following
empirical formula (M. J. Holman and P. A. Wiegert 1999),
which is valid for coplanar and initially circular planetary
orbits:
\begin{equation}
a_c=(0.464-0.380M_b-0.631e+0.586M_be+0.15e^2-0.198M_be^2)a.
\end{equation}
The results can be found in Table 9. It is noted that the stability
borders around the individual stars of each system are quite similar. 
This is due to the fact that all our systems consist of
stars with similar masses.

\begin{figure}
\centering
\includegraphics[width=120mm,height=60mm]{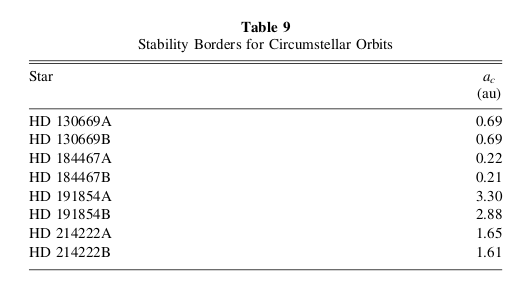}
\end{figure}

\subsection{Habitability}
The discovery of Earth-sized exoplanets has ignited a discussion
about the potential of finding life outside our solar system.
Determining whether a planet may be habitable is a complex and
multifaceted matter (e.g., M. J. Way and N. Georgakarakos 2017;
V. S. Meadows and R. K. Barnes 2018; R. Schwarz et al. 2018;
M. J. Way et al. 2023). We saw in the previous section that it is
possible for planetary bodies to be on stable orbits around the four
binary systems of this work. Orbital stability is usually one of the
first requirements when looking for potentially habitable planets.
A planet needs to be on an orbit that provides a suitable
environment for the development of life. Strong gravitational
interactions between the stellar binary and a planet can lead to
chaotic orbital evolution, which eventually may lead to the
ejection of the planet from the system. Therefore, the investigation
of the long-term dynamical stability of the planetary orbit is a
fundamental requirement when looking for habitable worlds.

We are now going to explore the potential for each one of our
systems to host a planet within the respective habitable zone.
The habitable zone is the area around a star where a planet with
an Earth-like atmosphere and on a circular orbit can retain liquid
water on its surface (J. F. Kasting et al. 1993). Planetary
systems, however, often consist of more than two bodies
(multistar systems or single-star multiplanet systems). This
implies that even if the planet is on an initially circular orbit, the
latter will evolve into an elliptical one, meaning that the distance
between the stars and the planets will vary on different
timescales (e.g., N. Georgakarakos 2003; N. Georgakarakos and
S. Eggl 2015). This means the classical (or static) habitable
zone, as defined previously, may not be an accurate marker for
making an initial assessment of the potential habitability status
of a planet. In a series of papers, we have developed an
analytical method to calculate planetary habitable zones in
binary systems or systems with more than one planet, taking
into consideration the orbital evolution of our potentially
habitable world (S. Eggl et al. 2012; N. Georgakarakos et al.
2018, 2021). A nice review on that matter can be found in
S. Eggl et al. (2020).

In the context of that work we defined three types of so-
called dynamically informed habitable zones (DIHZ): the
permanently habitable zone (PHZ), which is the area around
the star(s) where the planet always stays within habitable
radiation limits; the averaged habitable zone (AHZ), where the
planet is on average within habitable radiation limits; and
finally the extended habitable zone (EHZ), where the planet is
on average within habitable radiation limits plus or minus one
standard deviation. We apply our method to all four binary
systems. For circumbinary habitable zones we used the method
presented by N. Georgakarakos et al. (2021), while for
exploring habitable zones around one of the stars of the
binary system, we used results from S. Eggl et al. (2012). A
graphical representation of the results can be found in Figure 7.
In the calculations that led to these plots, we have allowed the
binary eccentricity to vary to observe the effect of that
parameter on the DIHZs.

Our calculations revealed that none of our systems have
circumbinary habitable zones. This is because the area close to
the stars is dynamically unstable; therefore, we cannot determine any habitable zone. The same happens with HD 184467 when looking for circumstellar habitable zones. As we
saw from our stability analysis, a planet would need to be very
close to either star to survive the gravitational pull of the other
one, and therefore, as in the circumbinary scenario, the
habitable zones of this system lie in dynamically unstable
areas.

On the other hand, as seen in Figure 7, HD 130669, HD
191854, and HD 214222 allow the existence of habitable zones
around their stars. Although HD 130669 shows good potential
for having habitable zones around its stars, dynamical
instability starts to settle into the habitable zone as the binary
eccentricity increases. As a result, when the eccentricity
reaches the current value of the system, the whole range of the
habitable zone is dynamically unstable. The situation is more
promising for HD 191854, where even for high values of the
binary eccentricity, the PHZ covers almost the whole extent of
the habitable zone. This happens because the system is the
widest binary among all four systems under consideration.
Finally, HD 214222 stands between the other two systems in
terms of the existence of habitable zones around its stellar
components, as seen in the bottom row of Figure 7.

\begin{figure}
\begin{center}
\includegraphics[width=70mm]{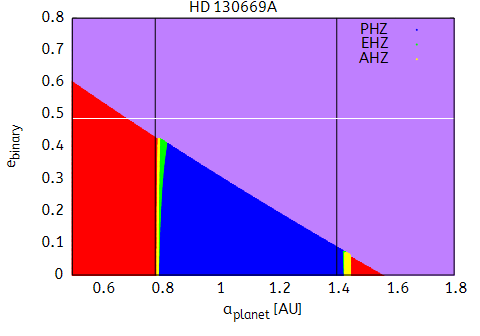}
\includegraphics[width=70mm]{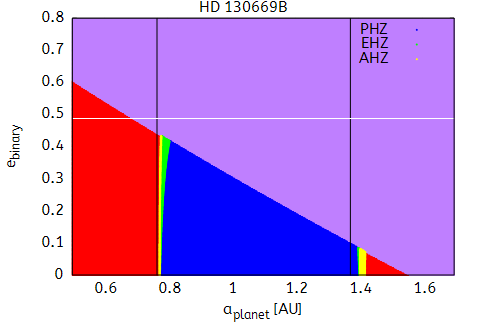}
\includegraphics[width=70mm]{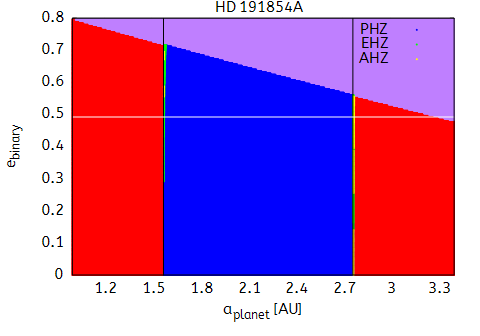}
\includegraphics[width=70mm]{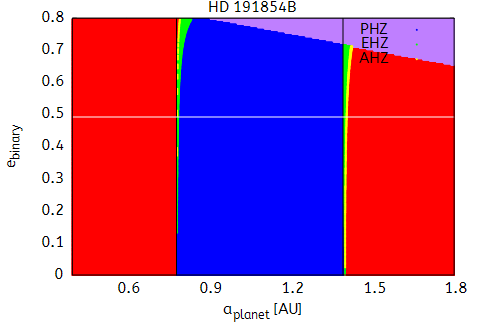}
\includegraphics[width=70mm]{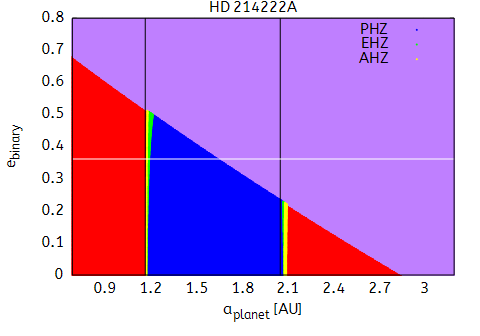}
\includegraphics[width=70mm]{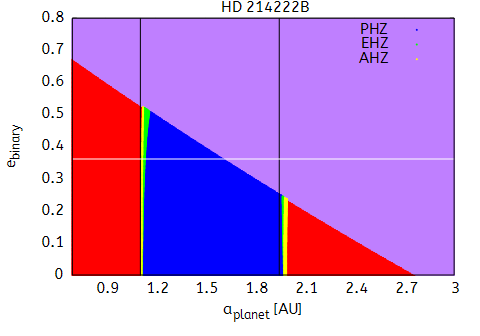}
\caption{Circumstellar habitable zones for HD 130669, HD191854, and HD214222. Blue color: PHZ, yellow color: AHZ, green color: EHZ, red color: uninhabitable area, purple color: areas of dynamical instability.  The white line marks the eccentricity of the stellar binary, while the vertical black lines are the borders of the classical habitable zone.}
\end{center}
\end{figure}

\section{Concluding Remarks}

The present work showcases the effectiveness of integrating
precise spectroscopic orbits with high-quality visual observations to obtain 
an all-inclusive set of orbital and physical
parameters for four double-lined spectroscopic binaries: HD
130669, HD 184467, HD 191854, and HD 214222. By refining
the orbital parallax, individual masses, and orbital separations,
this approach resolves earlier discrepancies found in Hipparcos
and Gaia (DR1/DR2/EDR3) measurements and results in
more accurate distance and mass estimates.

By applying Al-Wardat’s spectrophotometric modeling
technique, it was possible to estimate the essential stellar
properties of each component, such as temperatures, luminos-
ities, and radii. This information then facilitated an evaluation
of whether these systems could support stable, habitable
environments for any potential orbiting planets. Figure 8
presents all the previous steps and summarizes what we have
accomplished in this work. In particular,\\
1. Accurate masses and orbital parallaxes. Combining the
spectroscopic mass ratio with top-tier visual orbits
produced well-defined values for the total and individual
stellar masses. This integrated method also allowed for
an independent calculation of the orbital parallax, which
generally aligns well with and refines both Hipparcos and
Gaia data.\\
2. Consistency of mass ratios. The calculated and spectro-
scopic mass ratios agreed closely for HD 130669 and HD
184467, and were similarly consistent for HD 214222.
On the other hand, HD 191854 displayed larger
uncertainties, suggesting that further observations may
be needed to reduce measurement errors.\\
3. Stable orbital configurations. Preliminary stability ana-
lyses indicate that some of these binary systems could
support circumstellar or circumbinary planets, depending
on orbital size and eccentricity. While stable orbits are
feasible for specific configurations, higher eccentricities
can limit or undermine the habitable zone.\\
4. Habitability potential. For systems with intermediate
eccentricities and sufficiently separated stellar compo-
nents, there may be regions where terrestrial planets could 
maintain liquid water at the surface. Even so, the
specific range of habitability will hinge on both orbital
geometry and stellar properties.

Overall, this study underscores the importance of integrating
multiple observation techniques — spectroscopic, astrometric,
and interferometric — to attain accurate orbits and stellar
properties in multiple-star systems. Such high-precision data
are essential for understanding planet formation and the
likelihood of habitable conditions around Sun-like stars in
binary environments.

Figure 8 summarizes the methodological workflow applied
in this study, from data acquisition and orbital modeling to the
final habitability and dynamical stability assessments. By
combining spectroscopic and visual analysis of SB2 systems,
we can determine precise physical and orbital parameters that
are necessary for assessing dynamical stability and habit-
ability. Spectroscopic and calculated mass ratios are in good
agreement, validating the robustness of the methods used and
demonstrating the importance of multitechnique approaches. It
will also be possible to apply this methodology to larger
samples of SB2 systems once Gaia DR4 is released, which will
increase astrometric precision and orbital accuracy. Incorporating 
more targets and extending observational baselines will
increase the statistical power of habitability assessments.
Further interferometric observations and high-resolution
spectroscopic observations will help refine orbital inclinations
and resolve small-angle distance systems. It will be possible to
gain a deeper understanding of planet-hosting potential in
binary environments as a result of these developments.

\begin{figure}
\centering
\includegraphics[width=130mm,height=130mm]{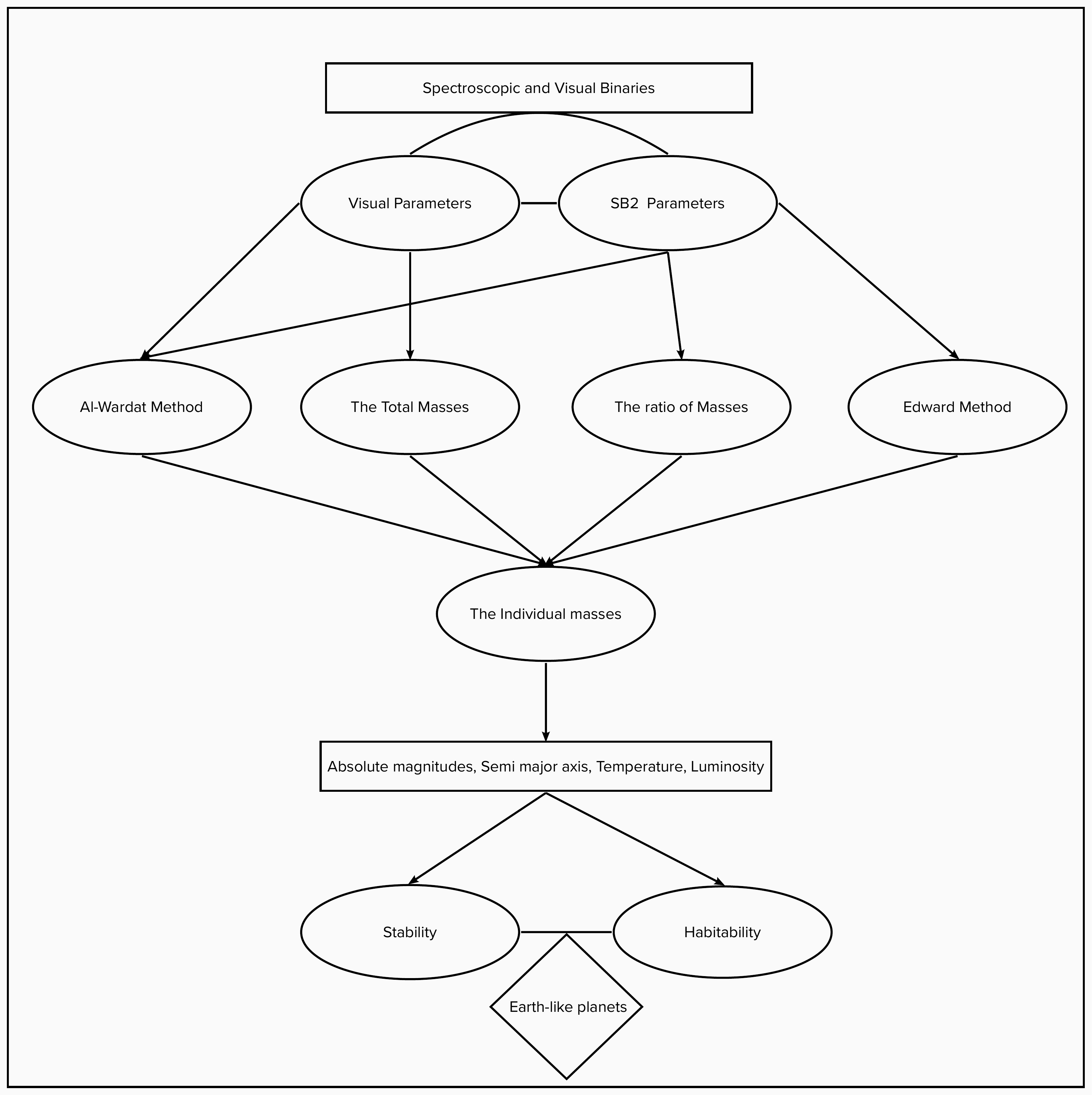}
\caption{A flowchart of the methodology used in this work.}
\end{figure}

\section*{Acknowledgments}
We are grateful to the US Naval Observatory for both the
Fourth Catalog of Interferometric Observations of Binary Stars
and the Washington Double Star Catalog. Also to Centre de
données astronomiques de Strasbourg (CDS), Strasbourg,
France, for the available SIMBAD database. This work has
made use of data from the European Space Agency (ESA)
mission Gaia\\
(https://www.cosmos.esa.int/gaia), processed
by the Gaia Data Processing and Analysis Consortium\\ 
(DPAC;https://www.cosmos.esa.int/web/gaia/dpac/consortium). It
also has made use of SAO/NASA, the SIMBAD database,
the Fourth Catalog of Interferometric Measurements of Binary
Stars, IPAC data systems, and codes of Al-Wardat’s method
for analyzing binary and multiple stellar systems.

\section*{ORCID iDs}
Ahmad Abushattalaa https://orcid.org/0000-0002-7796-6562\\
Nikolaos Georgakarakosaa https://orcid.org/0000-0002-
7071-5437\\
Mashhoor A. Al-Wardataa https://orcid.org/0000-0002-
1422-211X\\
Hassan B. Haboubiaa https://orcid.org/0009-0004-7254-0724\\
Deshinta Arrova Dewiaa https://orcid.org/0000-0003-1488-7696\\
Enas M. Abu-Alrobaa https://orcid.org/0000-0001-7718-9209\\
Abdallah M. Husseinaa https://orcid.org/0000-0002-
0738-4305\\

\section*{References}
Abushattal, A., Al-Wardat, M., Taani, A., Khassawneh, A., and Al-Naimiy, H. 2019, JPCS, 1258, 012018\\
Abushattal, A., Alrawashdeh, A., and Kraishan, A. 2022a, CoBAO, 69, 251\\
Abushattal, A., Kraishan, A., and Alshamaseen, O. 2022b, CoBAO, 69, 235\\
Abushattal, A. A., Al-Wardat, M. A., Horch, E. P., et al. 2024, AdSpR,
73, 1170\\
Abushattal, A. A., Docobo, J. A., and Campo, P. P. 2019b, AJ, 159, 28\\
Abushattal, A. A. M. 2017, PhD thesis, Universidade de Santiago de
Compostela\\
Al-Tawalbeh, Y. M., Hussein, A. M., Taani, A. A., et al. 2021, AstBu, 76, 71\\
Al-Wardat, M. 2002a, BSAO, 53, 58\\
Al-Wardat, M. 2002b, ESASP, 53, 51\\
Al-Wardat, M. 2007, AN, 328, 63\\
Al-Wardat, M. 2008, AstBu, 63, 361\\
Al-Wardat, M. 2009, AN, 330, 385\\
Al-Wardat, M. 2012, PASA, 29, 523\\
Al-Wardat, M., Balega, Y. Y., Leushin, V., et al. 2014a, AstBu, 69, 58\\
Al-Wardat, M., Balega, Y. Y., Leushin, V., et al. 2014b, AstBu, 69, 198\\
Al-Wardat, M., Docobo, J., Abushattal, A., and Campo, P. 2017a, AstBu, 72, 24\\
Al-Wardat, M., Docobo, J., Abushattal, A., and Campo, P. 2017b, AstBu, 72, 24\\
Al-Wardat, M., and Widyan, H. 2009, AstBu, 64, 365\\
Al-Wardat, M. A. 2002, BSAO, 53, 51\\
Al-Wardat, M. A. 2007, AN, 328, 63\\
Al-Wardat, M. A. 2009, AN, 330, 385\\
Al-Wardat, M. A. 2012, PASA, 29, 523\\
Al-Wardat, M. A., El-Mahameed, M. H., Yusuf, N. A., Khasawneh, A. M., and
Masda, S. G. 2016, RAA, 16, 166\\
Al-Wardat, M. A., Hussein, A. M., Al-Naimiy, H. M., and Barstow, M. A.
2021a, PASA, 38, e002\\
Al-Wardat, M. A., Abu-Alrob, E., Hussein, A. M., et al. 2021b, RAA, 21, 161\\
Alameryeen, H., Abushattal, A., and Kraishan, A. 2022, CoBAO, 69, 242\\
Algnamat, B., Abushattal, A., Kraishan, A., and Alnaimat, M. 2022, CoBAO,
69, 223\\
Alnaimat, S., Jameel, R., Abushattal, A., and Al-Wardat, M. 2025, in Proc. 14th
Arabic Conf. Arab Union for Astronomy and Space Sciences: AUASS-CONF23, Vol. 420 (Berlin: Springer Nature)\\
Armstrong, J. T., Clark III, J. H., Gilbreath, G. C., et al. 2004, Proc. SPIE,
5491, 1700\\
Balega, I., Bonneau, D., and Foy, R. 1984, A$\&$AS, 57, 31\\
Bonneau, D., Balega, Y., Blazit, A., et al. 1986, A$\&$AS, 65, 27\\
Borucki, W. J. 2016, RPPh, 79, 036901\\
Brown, A. G., Vallenari, A., Prusti, T., et al. 2016, A$\&$A, 595, A2\\
Christy, J. W., and Walker, R. 1969, PASP, 643\\
Docobo, J., Balega, Y., Campo, P., and Abushattal, A. 2018, WDS, 196, 1\\
Docobo, J., Balega, Y., Ling, J., Tamazian, V., and Vasyuk, V. 2000, AJ,
119, 2422\\
Docobo, J., Campo, P., Gomez, J., and Horch, E. P. 2018c, AJ, 156, 185\\
Docobo, J., Elipe, A., and Abad, A. 1988, Ap$\&$SS, 142, 195\\
Docobo, J. A. 1985, CeMec, 36, 143\\
Docobo, J. A. 2012, in Orbital Couples: Pas de Deux in the Solar System and
the Milky Way, ed. F. Arenou and D. Hestroffer (Paris: Observatoire de
Paris), 119\\
Docobo, J. A., Campo, P. P., Andrade, M., and Horch, E. P. 2014, AstBu,
69, 461\\
Docobo, J. A., Griffin, R. F., Campo, P. P., and Abushattal, A. A. 2017,
MNRAS, 469, 1096\\
Drimmel, R., Bucciarelli, B., and Inno, L. 2019, RNAAS, 3, 79\\
Duchêne, G., and Kraus, A. 2013, ARA$\&$A, 51, 269\\
Duquennoy, A., and Mayor, M. 1991, A$\&$A, 248, 485\\
Edwards, T. 1976, AJ, 81, 245\\
Eggen, O. 1965, AJ, 70, 19\\
Eggen, O. J. 1955, PASP, 67, 169\\
Eggl, S., Georgakarakos, N., and Pilat-Lohinger, E. 2020, Galax, 8, 65\\
Eggl, S., Pilat-Lohinger, E., Georgakarakos, N., Gyergyovits, M., and Funk, B.
2012, ApJ, 752, 74\\
Esa 1997, yCat, I/239, 323, L49–L52\\
Fardal, M. A., van der Marel, R., del Pino, A., and Sohn, S. T. 2021, AJ, 161, 58\\
Farrington, C. D., Ten Brummelaar, T., Mason, B., et al. 2010, AJ, 139, 2308\\
Fontanive, C., and Bardalez Gagliuffi, D. 2021, FrASS, 8, 16\\
Gaia Collaboration, Brown, A. G. A., Vallenari, A., et al. 2021, A$\&$A, 649, A1\\
Gaia Collaboration, et al. 2018a, yCat, 1345, 0\\
Gaia Collaboration, Mignard, F., Klioner, S., et al. 2018b, A$\&$A, 616, A14\\
Gaspar, A., Rieke, G. H., and Ballering, N. 2016, ApJ, 826, 171\\
Georgakarakos, N. 2003, MNRAS, 345, 340\\
Georgakarakos, N., and Eggl, S. 2015, ApJ, 802, 94\\
Georgakarakos, N., and Eggl, S. 2019, MNRAS, 487, L58\\
Georgakarakos, N., Eggl, S., Ali-Dib, M., and Dobbs-Dixon, I. 2024, AJ,
168, 224\\
Georgakarakos, N., Eggl, S., and Dobbs-Dixon, I. 2018, ApJ, 856, 155\\
Georgakarakos, N., Eggl, S., and Dobbs-Dixon, I. 2021, FrASS, 8, 44\\
Girardi, L., Bressan, A., Bertelli, G., and Chiosi, C. 2000, A$\&$AS, 141, 371\\
Griffin, R. F. 2012, Obs, 132, 309\\
Griffin, R. F. 2015, Obs, 135, 321\\
Heintz, W. 1997, ApJS, 111, 335\\
Hog, E., Fabricius, C., Makarov, V. V., et al. 2000, Technical Report, Naval
Observatory, Washington, DC\\
Holman, M. J., and Wiegert, P. A. 1999, AJ, 117, 621\\
Horch, E. P., Veillette, D. R., Gallé, R. B., et al. 2009, AJ, 137, 5057\\
Horch, E. P., Van Altena, W. F., Demarque, P., et al. 2015, AJ, 149, 151\\
Howell, S. B., Scott, N. J., Matson, R. A., et al. 2021, FrASS, 8, 10\\
Hussein, A. M., Al-Wardat, M. A., Abushattal, A., et al. 2022, AJ, 163, 182\\
Jack, D., Hernández Huerta, M. A., and Schröder, K.-P. 2020, AN, 341, 616\\
Kasting, J. F., Whitmire, D. P., and Reynolds, R. T. 1993, Icar, 101, 108\\
Keenan, P. C., and McNeil, R. C. 1989, ApJS, 71, 245\\
Kiefer, F., Halbwachs, J.-L., Lebreton, Y., et al. 2018, MNRAS, 474, 731\\
Labeyrie, A. 1970, A$\&$A, 6, 85\\
Lester, K. V., Gies, D. R., Schaefer, G. H., et al. 2019a, AJ, 157, 140\\
Lester, K. V., Gies, D. R., Schaefer, G. H., et al. 2019b, AJ, 158, 218\\
Lester, K. V., Fekel, F. C., Muterspaugh, M., et al. 2020, AJ, 160, 58\\
Masda, S., Docobo, J., Hussein, A., et al. 2019a, AstBu, 74, 464\\
Masda, S. G., Al-Wardat, M. A., Neuhäuser, R., and Al-Naimiy, H. M. 2016,
RAA, 16, 112\\
Masda, S. G., Al-Wardat, M. A., and Pathan, J. K. M. K. 2018, RAA, 18, 072\\
Masda, S. G., Al-Wardat, M. A., and Pathan, J. M. 2019b, RAA, 19, 105\\
Matson, R. A., Howell, S. B., Horch, E. P., and Everett, M. E. 2018, AJ, 156, 31\\
McAlister, H., Hendry, E., Hartkopf, W., Campbell, B., and Fekel, F. 1983,
ApJS, 51, 309\\
McAlister, H. A. 1976, PASP, 88, 957\\
McAlister, H. A. 1977, ApJ, 212, 459\\
McAlister, H. A. 1978, ApJ, 223, 526\\
Meadows, V. S., and Barnes, R. K. 2018, in Handbook of Exoplanets, ed.
H. J. Deeg and J. A. Belmonte, Vol. 57 (Berlin: Springer)\\
Michel, K.-U., and Mugrauer, M. 2021, FrASS, 8, 14\\
Mitrofanova, A., Dyachenko, V., Beskakotov, A., et al. 2021, AJ, 162, 156\\
Netopil, M. 2017, MNRAS, 469, 3042\\
Perryman, M., et al. 1997, star, 2, 2\\
Piccotti, L., Docobo, J. Á., Carini, R., et al. 2020, MNRAS, 492, 2709\\
Pilat-Lohinger, E., and Bazsó, Á. 2021, FrASS, 8, 47\\
Pourbaix, D. 2000, A$\&$AS, 145, 215\\
Pourbaix, D., Tokovinin, A. A., Batten, A. H., et al. 2004, A$\&$A, 424, 727\\
Prusti, T., De Bruijne, J., Brown, A. G., et al. 2016, A$\&$A, 595, A1\\
Raghavan, D., McAlister, H. A., Henry, T. J., et al. 2010, ApJS, 190, 1\\
Ren, S., and Fu, Y. 2010, AJ, 139, 1975\\
Ricker, G. R., Latham, D., Vanderspek, R., et al. 2010, AAS, 215, 450\\
Roman, N. G. 1955, ApJS, 2, 195\\
Schwarz, R., Bazsó, Á., Georgakarakos, N., et al. 2018, MNRAS, 480,
3595\\
Soubiran, C., Le Campion, J.-F., Brouillet, N., and Chemin, L. 2016, A$\&$A,
591, A118\\
Starikova, G. 1978, SvAL, 4, 296\\
Stassun, K. G., and Torres, G. 2018, ApJ, 862, 61\\
Taani, A., Abushattal, A., and Mardini, M. K. 2019, AN, 340, 847\\
Tokovinin, A. 1992a, International Astronomical Union Colloquium, Vol. 135
(Cambridge: Cambridge Univ. Press), 573\\
Tokovinin, A. 1992b, A$\&$A, 256, 121\\
Tokovinin, A. 2014, AJ, 147, 86\\
Tokovinin, A., and Horch, E. P. 2016, AJ, 152, 116\\
Tokovinin, A., Mason, B. D., Mendez, R. A., Horch, E. P., and Briceno, C.
2019, AJ, 158, 48\\
van Leeuwen, F. 2007, A$\&$A, 474, 653\\
Vasiliev, E. 2019, MNRAS, 489, 623\\
Wang, X.-L., Ren, S.-L., and Fu, Y.-N. 2016, RAA, 16, 033\\
Way, M. J., and Georgakarakos, N. 2017, ApJL, 835, L1\\
Way, M. J., Georgakarakos, N., and Clune, T. L. 2023, AJ, 166, 227\\
Wenger, M., Ochsenbein, F., Egret, D., et al. 2000, A$\&$AS, 143, 9\\
Yousef, Z. T., Annuar, A., Hussein, A. M., et al. 2021, RAA, 21, 114\\
Zinn, J. C., Huber, D., Pinsonneault, M. H., and Stello, D. 2017, ApJ, 844, 166

\end{document}